\documentclass[%
 aip,
 amsmath,amssymb,
 reprint,%
]{revtex4-1}

\usepackage{graphicx}
\usepackage{dcolumn}
\usepackage{bm}

\usepackage[utf8]{inputenc}
\usepackage[T1]{fontenc}
\usepackage{mathptmx}
\usepackage{hyperref}
\hypersetup{
    colorlinks   = true,
    citecolor    = blue,
    linkcolor    = blue}

\DeclareMathAlphabet{\mathcal}{OMS}{cmsy}{m}{n} 
\renewcommand{\v}[1]{\ensuremath{\bm{#1}}} 
\newcommand{\gv}[1]{\ensuremath{\mbox{\bm$ #1 $}}} 
\newcommand{\uv}[1]{\ensuremath{\bm{#1}}} 
\newcommand{\avg}[1]{\left< #1 \right>} 
\renewcommand{\d}[2]{\frac{d #1}{d #2}} 
\newcommand{\pd}[2]{\frac{\partial #1}{\partial #2}} 
 
\newcommand{\partialy}[1]{\partial_y #1} 
\newcommand{\grad}[1]{\gv{\nabla} #1} 
\renewcommand{\div}[1]{\gv{\nabla} \cdot #1} 
\newcommand{\curl}[1]{\gv{\nabla} \times #1} 
\let\baraccent=\= 
\renewcommand{\=}[1]{\stackrel{#1}{=}} 
\newcommand{\delpar}[1]{\nabla_{\parallel} #1} 
\newcommand{\gradperp}[1]{{{\gv{\nabla}}_{\perp}} {#1}} 
\newcommand{\divperp}[1]{{\gv{\nabla}}_{\perp}\cdot #1} 

\newcommand{\zhat}{\v{\hat{z}}}
\newcommand{\curv}[1]{{C}_{\left({#1}\right)}}
\newcommand{\ncurv}[1]{\mathcal{C}_{\left({#1}\right)}}

\newcommand{\txtS}[2]{#1_{\text{#2}}} 

\newcommand{\txtSb}[3]{#1_{#2\text{#3}}}
\newenvironment{eqnal}{\equation\aligned}{\endaligned\endequation}

\newcommand{\sind}{s}
\newcommand{\qs}{q_s}
\newcommand{\ms}{m_s}
\newcommand{\BO}{B_0}
\newcommand{\omegaci}{\Omega_i}
\newcommand{\omegace}{\Omega_e}
\newcommand{\omegacs}{\Omega_\sind}

\newcommand{\cseO}{c_{se0}}
\newcommand{\cse}{c_{se}}
\newcommand{\rhos}{\rho_{s}}
\newcommand{\bhat}{\uv{b}}
\newcommand{\cur}{j_{\parallel}}
\newcommand{\vpi}{u_{\parallel i}}						
\newcommand{\vpe}{u_{\parallel e}}					
\newcommand{\vvExB}{\v{v}_{E}}						
\newcommand{\vvds}{\v{v}_{ds}}

\newcommand{\kappacurv}{\gv{\kappa}}			
\newcommand{\vB}{\v{B}}								
\newcommand{\gvort}{\omega} 
\newcommand{\Tsrc}{\txtS{T}{src}}
\newcommand{\xSrc}{\txtS{R}{src}}
\newcommand{\sigSrc}{\txtS{\sigma}{src}}
\newcommand{\tpar}{\tau_\parallel}
\newcommand{\Lc}{L_c}
\newcommand{\Lz}{L_z}
\newcommand{\Ly}{L_y}
\newcommand{\Lx}{L_x}

\newcommand{\qpare}{q_{\parallel e}}
\newcommand{\qpari}{q_{\parallel i}}
\newcommand{\gammae}{\gamma_e}
\newcommand{\gammai}{\gamma_i}
\newcommand{\cs}{c_s}
\newcommand{\upars}{u_{\parallel s}}
\newcommand{\upari}{u_{\parallel i}}
\newcommand{\upare}{u_{\parallel e}}
\newcommand{\kappapari}{\kappa^i_\parallel}
\newcommand{\kappapare}{\kappa^e_\parallel}
\newcommand{\kpar}{k_{\parallel}}
\newcommand{\SOfl}{S_{0}^{\text{fl}}}

\newcommand{\jac}{\mathcal{J}}
\newcommand{\jacs}{\jac_\sind}
\newcommand{\vpar}{v_{\parallel}}
\newcommand{\fs}{f_\sind}

\newcommand{\vBS}{\vB^\star}
\newcommand{\BparS}{B^\star_\parallel}
\newcommand{\vBparS}{\vB^\star_\parallel}
\newcommand{\dmu}{\text{d}\mu}
\newcommand{\dvpar}{\text{d}\vpar}

\newcommand{\alphad}{\alpha_d}

\newcommand{\epa}{\epsilon_R}
\newcommand{\epv}{\epsilon_v}
\newcommand{\epgi}{\epsilon_{G_i}}
\newcommand{\epge}{\epsilon_{G_e}}

\newcommand{\etapar}{\eta_{\parallel}}
\newcommand{\kappai}{\kappa^i}
\newcommand{\kappae}{\kappa^e}

\newcommand{\tRef}{\txtS{t}{ref}}

\newcommand{\LparRef}{\txtSb{L}{\parallel}{ref}}
\newcommand{\BRef}{\txtS{B}{ref}}
\newcommand{\RRef}{\txtS{R}{ref}}
\newcommand{\nRef}{\txtS{n}{ref}}
\newcommand{\TeRef}{\txtSb{T}{e}{,ref}}
\newcommand{\TiRef}{\txtSb{T}{i}{,ref}}

\newcommand{\vpeRef}{\txtSb{u}{\parallel e}{,ref}}

\newcommand{\cseRef}{\txtSb{c}{se}{,ref}}
\newcommand{\taueRef}{\txtSb{\tau}{e}{,ref}}
\newcommand{\tauiRef}{\txtSb{\tau}{i}{,ref}}

\newcommand{\Nx}{N_x}

\newcommand{\Nz}{N_z}

\newcommand{\Dz}{\Delta z}

\newcommand{\gdb}{\texttt{GDB}}
\newcommand{\gkeyll}{\texttt{Gkeyll}}
\newcommand{\grillix}{\texttt{GRILLIX}}
\newcommand{\bout}{\texttt{BOUT++}}
\newcommand{\gbs}{\texttt{GBS}}
\newcommand{\tokam}{\texttt{TOKAM3X}}
\newcommand{\ExB}{$E\times B$}
\newcommand{\xgc}{\texttt{XGC1}}
\newcommand{\elmfire}{\texttt{ELMFIRE}}
\newcommand{\cogent}{\texttt{COGENT}}
\newcommand{\gysela}{\texttt{GYSELA}}
\newcommand{\gene}{\texttt{GENE}}
\newcommand{\picls}{\texttt{PICLS}}

\begin{document}

\preprint{AIP/123-QED}

\title{Fluid \& Gyrokinetic turbulence in open field-line, helical plasmas}

\author{M. Francisquez}
\affiliation{ 
MIT Plasma Science and Fusion Center, Cambridge, MA 02139}%

 \altaffiliation[Also at ]{Princeton Plasma Physics Laboratory.}
 \email{manafr@mit.edu}
\author{T. N. Bernard}%
\affiliation{General Atomics, PO Box 85608, San Diego, CA 92186, USA}%
\affiliation{Institute for Fusion Studies, University of Texas at Austin, Austin, TX 78712, USA}

\author{B. Zhu}
\affiliation{Lawrence Livermore National Laboratory, Livermore, CA 94550, USA}%

\author{A. Hakim}
\affiliation{Princeton Plasma Physics Laboratory, Princeton, NJ 08543, USA}%

\author{B. N. Rogers}
\affiliation{Dartmouth College, Hanover, NH 03755, USA}%

\author{G. W. Hammett}
\affiliation{Princeton Plasma Physics Laboratory, Princeton, NJ 08543, USA}%

\date{\today}

\begin{abstract}
Two-fluid Braginskii codes have simulated open-field line turbulence for over a decade, and only recently has it become possible to study these systems with continuum gyrokinetic codes. This work presents a first-of-its-kind comparison between fluid and (long-wavelength) gyrokinetic models in open field-lines, using the \gdb~and \gkeyll~codes to simulate interchange turbulence in the Helimak device at the University of Texas (T. N. Bernard, et. al., Phys. of Plasmas 26, 042301 (2019)). Partial agreement is attained in a number of diagnostic channels when the \gdb~sources and sheath boundary conditions (BCs) are selected carefully, especially the heat-flux BCs which can drastically alter the temperature. The radial profile of the fluctuation levels is qualitatively similar and quantitatively comparable on the low-field side, although statistics such as moments of the probability density function and the high-frequency spectrum show greater differences. This comparison indicates areas for future improvement in both simulations, such as sheath BCs, as well as improvements in \gdb~like particle conservation and spatially varying thermal conductivity, in order to achieve better fluid-gyrokinetic agreement and increase fidelity when simulating experiments.
\end{abstract}

\maketitle

\section{\label{sec:intro}Motivation and overview}

Helical magnetic field devices, such as the Helimak (University of Texas) and TORPEX (EPFL), provide a useful environment for refining our understanding of open field-line toroidal systems. These devices have important ingredients of tokamak scrape-off layer (SOL) turbulence: parallel transport onto sheath regions, turbulent cross-field transport, curvature and $\nabla B$ drifts, and interaction with plasma-facing materials, main chamber neutrals and radio-frequency (RF) sources. Numerous aspects of tokamak fusion plasma operation are highly dependent on the conditions in the SOL. The fusion performance of the core, for example, is thought to be directly dependent on the plasma temperature at the pedestal top~\cite{Kotschenreuther1996} which is dynamically affected by the properties of the SOL plasma. The SOL is the site of sometimes deleterious field-aligned motion of particles and heat towards the tiles of the vessel wall. These are complicated and, to some extent, mitigated by the cross-field transport spreading loads over a larger surface of the wall~\cite{Stangeby2000}. Cross-field transport is not always desirable however, since coherent structures~\cite{Myra2006} and edge-localized modes~\cite{Pitts2007} can also impact the walls and cause sputtering, penetration of impurities and plasma cooling, all of which undermine the performance of the core. Understanding these processes, and gaining the ability to predict and optimize them, are highly desirable for designing and operating future experiments.

One way to study and predict the time evolution of experiments' edge is through direct numerical simulation. This approach has been made possible with both fluid and kinetic models in open field-line turbulent systems thanks to the advent of high-performance computing (HPC). In the next few years HPC will take another major step as exascale supercomputers become available, which strategies for burning plasma research plan to leverage in order to guide our understanding of current experiments and help to optimize the design of future devices~\cite{NAP2019}. A number of fluid and kinetic codes are currently being developed and upgraded to take advantage of these capabilities and deliver a realistic numerical description of laboratory plasmas.

Full-$f$ fluid codes, those which do not separate the evolution of equilibrium and fluctuating contributions to the plasma parameters, have modeled helical open-field line turbulence for over a decade~\cite{Ricci2008,Li2009,Ricci2010,Li2011}. These studies have consisted of solving a set of partial differential equations obtained from the drift-reduction of the Braginskii or Mikhailovskii fluid equations. A number of assumptions employed then have been done away with in modern Braginskii solvers, such a relaxing the Boussinesq approximation, using realistic and spatially-varying transport coefficients, including electromagnetic fluctuations and more complex geometries. Among the notable things uncovered by these studies are the appearance of large sheared velocity flows that quench turbulence when sources or field-line connection lengths are increased~\cite{Ricci2008}, reminiscent of the L-H transition in tokamaks. An early 2D solution of a simplified Braginskii model was also compared against experimental data, showing good agreement in several channels such as the global density and electron temperature profiles, and the frequency spectra~\cite{Li2009}. In global (not field-aligned) 3D simulations, it was also possible to discern the transition from interchange to drift-wave turbulence as one lowered the pitch angle (increased connection length) and lowered the collisionality~\cite{Ricci2010}.

As the collisionality decreases (or the temperature increases) the use of the Braginskii equations is often put into question, since these employ a short mean-free-path collisional closure to the system of moments of the kinetic equation. Despite this limitation, a variety of Braginskii codes have been developed recently to study the turbulence in the hot boundary plasma of tokamaks. A great effort is underway to improve the accuracy and robustness of codes such as \tokam~\cite{Tamain2016}, \gbs~\cite{Halpern2016}, \grillix~\cite{Stegmeir2018}, \bout~\cite{Dudson2009} and \gdb~\cite{Zhu2018}. Despite their use of collisional fluid equations in less collisional environments, several comparisons between their simulations and experiments have yielded satisfactory agreement~\cite{Halpern2015,Chen2017}. The reduced computational cost of fluid simulations also offer the ability to perform more parameter scans and iterative numerical simulation, which is often necessary to uncover the underlying physics. Perhaps for this reason alone there may always be an interest in fluid modeling, even if only as a step prior to kinetic simulation.

Yet the possibility remains that collisionless and other kinetic effects play crucial roles in the dynamics of boundary plasmas, and that these processes cannot be captured by Braginskii fluid codes. To address such concern several teams are also developing fluid models that are not derived under the assumption of strong collisionality~\cite{Waltz2019,Ng2018}. Extensive work is also being done in developing a new generation of solvers for the 5D gyrokinetic equation, a version of the Boltzmann kinetic equation averaged over the fast gyro-motion of particles around the magnetic field. Particle-in-cell (PIC) methods have accomplished a solution of this equation in both open and closed field lines; the \xgc~code, for example, has made valuable contributions to the prediction of heat-flux loads in current and future devices~\cite{Chang2017}. There is interest in cross-validating \xgc~results and also improving on its description of laboratory plasmas, for which other gyrokinetic codes are being developed, including~\gene~\cite{Pan2018}, \gysela~\cite{Grandgirard2016}, \elmfire~\cite{Chone2018}, \picls~\cite{Boesl2019} and \cogent~\cite{Dorf2016}. Among continuum codes, \gkeyll~pioneered the simulation of gyrokinetic turbulence in open field lines~\cite{Shi2017}. This approach was later used to study the SOL of the National Spherical Torus Experiment (NSTX)~\cite{Shi2019} and was incorporated in \gene~to model the Large Plasma Device (LAPD)~\cite{Pan2018}.

Currently no single code has all the ingredients required for a high degree of numerical realism. By comparing these different tools, we can learn which physics are exclusively kinetic and not captured by fluid frameworks, which parameter regimes can be safely studied with fluid models, and how one description can inform the improvement of the other. There is also interest in obtaining evidence of when certain numerical or analytical simplifications make no discernible difference, or when a given theoretical assumption proves too risky.

To this end, the Helimak device serves as a helpful testbed for the description of collisional, open field-line toroidal plasmas with both fluid and gyrokinetic models. Additionally, despite the simplified geometry and relatively high collisionality of the Helimak, predictive capability is still unattained and interesting open questions remain. We thus compared simulations of this system with both the \gdb~two-fluid code and the \gkeyll~gyrokinetic code. Although more sophisticated simulations of the Helimak are currently possible with these tools, we choose to compare the first published gyrokinetic simulations of Helimak~\cite{Bernard2019}, with a version of \gdb~that incorporates some simplifications commonly used by other Braginskii codes. In section~\ref{sec:models} we describe the fluid and gyrokinetic models, and some of the numerics employed to solve them. We direct the reader to other publications for additional details on the numerical methods of \gdb~\cite{Zhu2018} and \gkeyll~\cite{Shi2017a,Hakim2019}. The results of the fluid and gyrokinetic simulations are presented and analyzed in section~\ref{sec:results}, and we offer additional discussion and conclusions in section~\ref{sec:conclusion}.

\section{Description of gyrokinetic and fluid models} \label{sec:models}

\begin{figure}
\includegraphics[width=0.4\textwidth]{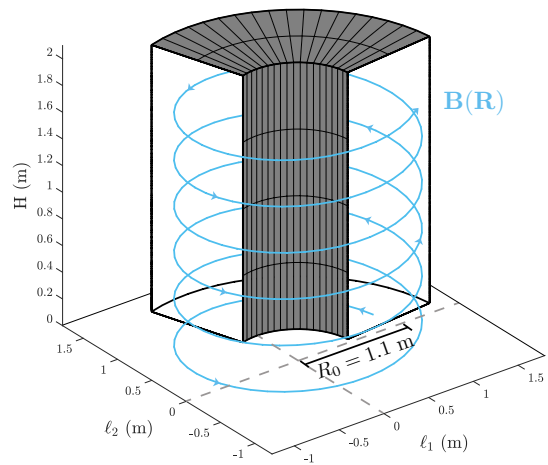}
\caption{\label{fig:Helimak} Sketch of the Texas Helimak geometry. Field lines terminate on bottom and top plates, where sheath regions are located.}
\end{figure}

In this section we present the gyrokinetic and fluid models used for this study and summarize their numerical aspects. For further details on the numerical implementation we refer the reader to the documentation on \gdb~\cite{Zhu2018} and \gkeyll~\cite{Shi2017a}. Both codes have been used to study the Texas Helimak toroidal device consisting of a $H=2$ m tall vessel with a rectangular cross section 1 m wide (see figure~\ref{fig:Helimak}). The major radius in the center of the plasma is $R_0=1.1$ m. The background helical magnetic field $B=B(R)$, composed of a toroidal $B_t$ and a vertical $B_v$ component, starts at the bottom plate and winds counterclockwise (as seen from above) until reaching the top of the vessel. We consider experiments with large pitch angle ($\propto B_v/B_t$) in which interchange modes with $\kpar\simeq 0$ dominate. Thus, as one winds around the machine once (following a field line) and displaces vertically by $\Ly=2\pi R B_v/B_t$, there will be little change in the plasma parameters. We can thus expect periodicity after every vertical segment $\Ly$ long.

\gkeyll~and \gdb~have been set up with field-aligned coordinate systems, meaning $(x,y,z)$ correspond to the radial ($R$), binormal and field-aligned directions. The computational domain corresponds to a flux tube that begins at the bottom of the device and after $N=H/\Ly$ turns ends at the top. The connection length is thus $\Lc=2\pi R N$. Both computational domains span the radial width of the vessel ($x\in[0.6~\text{m}, 1.6~\text{m}]$), the entire connection length ($z\in[-\Lc/2,\Lc/2]$), and have a restricted periodic binormal extent ($y\in[-\Ly/2,\Ly/2]$). Further explanation of the computational geometry can be found in a previous publication on \gkeyll~simulations of this machine~\cite{Bernard2019}.

We focus on an Argon case ($m_i/m_e=7.33\times10^4$) but using the reduced mass ratio $m_i/m_e=400$, with the magnetic field magnitude $B(R=R_0)=\BO=0.1$ T and a connection length of $\Lc=40$ m. Given the counter-clockwise rotation of the field, the unit vector along the background magnetic field is opposite the $z$-direction: $\bhat=\v{B}/B=-\zhat$. We neglect the shear in the magnetic field produced by the fact that $B_t\propto R^{-1}$ while $B_v$ is constant. Other plasma parameters are close to previous experiments and numerical simulations, with the density $n_{e0}=10^{16}$ m$^{-3}$ and the electron temperature $T_{e0}=10$ eV. Ions do not have time to thermalize with the electrons given the rapid charge-exchange and parallel losses, so we assume low temperature ions with a  source temperature of $T_{i0}=1$ eV.

\subsection{\gkeyll's gyrokinetic model}

We reproduce the description of the gyrokinetic model~\cite{Bernard2019} here for completeness and to motivate the choice of the sources in the fluid model. We are concerned with the electrostatic, long-wavelength limit (no Larmor-radius effects) of the full-$f$ gyrokinetic equation for the gyrocenter distribution function $\fs(\v{x},\vpar,\mu,t)$. This entails the continuum kinetic equation for species $\sind$
\begin{equation} \label{eq:fDot}
\pd{\jacs\fs}{t}+\div{\jacs\dot{\v{x}}\fs}+\pd{}{\vpar}\jacs\dot{\vpar}\fs = \jacs C[\fs]+\jacs S_\sind,
\end{equation}
where $C[\fs]$ incorporates the effects of collisions via the Dougherty operator
\begin{eqnal}
C[\fs] = \sum_{r}\nu_{sr}&\left\lbrace \pd{}{\vpar}\left[\left(\vpar-u_{\parallel sr}\right)\fs+v_{tsr}^2\pd{\fs}{\vpar}\right] \right. \\
&\left.\quad+\pd{}{\mu}\left[2\mu\fs+2\frac{m_sv_{tsr}^2}{B}\mu\pd{\fs}{\mu}\right]\right\rbrace
\end{eqnal}
and $S_\sind$ is a source of particles and energy. For like-particle collisions $u_{\parallel sr}=\upars$ and $v_{tsr}^2=v_{ts}^2=T_s/m_s$. Ion-electron collisions are neglected and electron-ion collisions use $u_{\parallel ei}=u_{\parallel i}$ and $v_{tei}^2=v_{te}^2+\left(u_{\parallel i}-u_{\parallel e}\right)^2/3$. The Jacobian of the coordinate transformation is $\jac=\BparS$ with $\vBparS=\vB+(B\vpar/\omegacs)\curl{\bhat}$ and $\BparS=\bhat\cdot\vBparS$, where $\omegacs$ is the gyrofrequency of species $s$, and we simply set $\vBparS\simeq\vB$. Given the Poisson bracket for this Hamiltonian system
\begin{equation}
\left\lbrace F,G \right\rbrace = \frac{\vBS}{\ms\BparS}\cdot\left(\grad{F}\pd{G}{\vpar}-\pd{F}{\vpar}\grad{G}\right)-\frac{1}{\qs\BparS}\bhat\cdot\grad{F}\times\grad{G},
\end{equation}
the advection velocities in phase-space are $\dot{\v{x}}=\{\v{x},H\}$ and $\dot{\vpar}=\{\vpar,H\}$, where the gyrocenter Hamiltonian is
\begin{equation}
H_\sind = \frac{1}{2}\ms\vpar^2+\mu B+\qs\phi.
\end{equation}
There are no Larmor-radius effects so we use $\phi$ in the Hamiltonian instead of the gyroaveraged potential $\avg{\phi}_\alpha$. This gyrokinetic system is closed by the long-wavelength gyrokinetic Poisson equation to compute the electrostatic potential:
\begin{equation} \label{eq:poisson}
    -\divperp{\frac{n_{i0}^gq_i^2\rho_{s0}^2}{T_{e0}}\gradperp{\phi}} = q_in_i^g(\v{x},t)-en_e(\v{x},t),
\end{equation}
with the ion sound gyro-radius $\rho_{s0}=c_{se0}/\omegaci$ given in terms of the zeroth-order ion sound speed $c_{se0}=\sqrt{T_{e0}/m_i}$ at the reference temperature $T_{e0}$. The guiding-center density $n_i^g$ is the zeroth velocity moment of the ion guiding center distribution function, $f_i$. Note that the ion guiding-center density on the left side of equation~\ref{eq:poisson} is taken to be the spatially constant, reference density ($n_{i0}^g=n_{e0}$). Similarly the variation of the magnetic field is not accounted for in the Poisson equation. This is akin to the Boussinesq approximation commonly made in Braginskii solvers.

The kinetic plasma model included the simplified phase-space source given by
\begin{equation} \label{eq:src}
S_\sind(\v{x},\vpar,\mu)=S_0\exp\left[-\frac{\left(R-\xSrc\right)^2}{2\sigSrc^2}\right]F_M(\vpar,\mu,\Tsrc).
\end{equation}
The velocity-space variation is given by the zero-flow normalized Maxwellian with temperature $\Tsrc$, $F_M(\vpar,\mu,\Tsrc)$. The radial location and width of the source are given by $\xSrc$ and $\sigSrc$, respectively. In the experiment the heating is provided by radio-frequency (RF) systems that interact with the electrons at the electron-cyclotron and the upper-hybrid resonances, primarily. Modeling this power source is complicated by the fact that the resonance location is dependent on the time-evolving plasma parameters. A practice of locating $S_\sind$ at a mean location of this resonance is followed here ($\xSrc=1.0$ m), and the width is chosen to be small to simulate the narrow RF source ($\sigSrc=0.01$ m). The absorption layer is broadened by turbulence. The appropriate amplitude of the source, and also the steady-state plasma profiles, were estimated with a 1D transport model~\cite{Shi2019} assuming a balance of the particle source (for a species $\sind$)
\begin{equation} \label{eq:nSrc}
S_{n,\sind}(\v{x}) = \frac{2\pi B}{\ms}\int\dvpar\dmu\thinspace S_s(\v{x},\vpar,\mu)
\end{equation}
and the parallel loss rate, $n_(\v{x},z)/\tpar$, with the parallel transit time defined as $\tpar=\Lc/(2c_s)$. The result is the approximate steady-state profile
\begin{equation} \label{eq:nIC}
n(x,z) = n_p\exp\left[-\frac{\left(R-\xSrc\right)^2}{2\sigSrc^2}\right]\frac{1+\sqrt{1-z^2/(\Lc/2)^2}}{2}
\end{equation}
The value of
$n_p=4.48\times10^{17}$ m$^{-3}$ 
was set such that the volume average of $n(x,z)$ is equal to $n_{e0}$. In order to maintain this profile with the source in equation~\ref{eq:nSrc} one can show that the amplitude of the source in equation~\ref{eq:src} must be $S_0\approx9.77\times10^{19}$ m$^{-3}$ s$^{-1}$, but \gkeyll~simulations were instead carried out with $S_0\approx 12.98\cdot(4{\times}10^{19}\text{ m}^{-3})\sqrt{(5/3)(T_e+T_i)/m_i}/\Lc = 8.6{\times} 10^{19}$ m$^{-3}$ s$^{-1}$. The temperature of the source's Maxwellian, $\Tsrc$, was also informed by this 1D transport calculation, which did not include parallel heat conduction. This heat transport is significant, which is why a higher value of $\Tsrc=(10/3)T_{e0}$ was employed. Since the source in equation~\ref{eq:src} has a non-drifting distribution in velocity space there is no net external addition of momentum, but there is an injection of energy given by
\begin{equation} \label{eq:enerSrc}
S_{E,\sind}(\v{x}) = \frac{2\pi B}{\ms}\int\dvpar\dmu\thinspace \left(\frac{1}{2}\ms\vpar^2+\mu B\right)S_s(\v{x},\vpar,\mu).
\end{equation}

\begin{figure}
\includegraphics[width=0.48\textwidth]{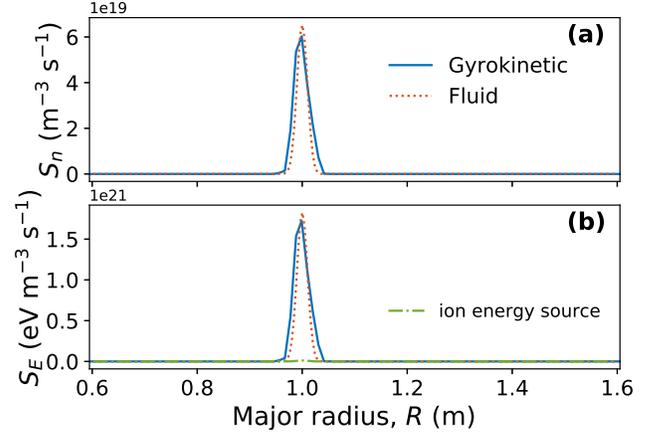}
\caption{\label{fig:Srcs} Particle (a) and energy (b) sources used in the gyrokinetic and fluid simulations. Sources had no variation in $y$ and $z$.}
\end{figure}

This gyrokinetic model is discretized with high-order discontinuous Galerkin (DG) schemes. Such approach can offer increased accuracy at a reduced cost compared to other numerical techniques, can be made to adapt to complex geometries, and improves data locality which is attractive for high-performance computing. Explicit third-order Runge-Kutta (SSP-RK3) time stepping was used. In this work the discrete, piecewise-linear ($p=1$) DG initial conditions and sources are obtained by evaluating their analytic function at the cell boundary nodes and using linear interpolation between them (more accurate quadrature methods are also available within~\gkeyll). The sources at $y=z=0$, for example, are shown in figure~\ref{fig:Srcs}. These figures are obtained by subdividing the $x$ domain into $\Nx(p+1)$ cells and plotting the cell-center value of the source ($\Nx$ is the number of cells along $x$).

The boundary conditions (BCs) on the distribution function $f_s$ are zero-flux along $x$ and periodic in $y$. The former is consistent with a homogeneous Dirichlet condition on $\phi$, which eliminates radial flows out of the domain. In the $z$ direction a model for
conducting sheath BCs with normally incident magnetic field lines are used~\cite{Shi2017,Shi2017a} (oblique incidence may be an interesting research topic to be examined in the future~\cite{Geraldini2017}). The conducting sheath is produced by solving for the potential at the sheath entrance, $\phi_{sh}=\phi(z=\pm\Lc/2)$, with the Poisson equation~\ref{eq:poisson}. Electrons with velocities $\vpar>\sqrt{2e\phi_{sh}/m_e}$ are lost through the sheath, while those with velocity lower than this but directed towards the sheath are reflected. The ions are allowed to pass through the sheath and become absorbed at whatever velocities they are accelerated to by the potential. We simply require that there are no incoming ions from the sheath, i.e. $f_i(x,y,z=-\Lc/2,\vpar,\mu)=0$ and $f_i(x,y,z=+\Lc/2,\vpar,\mu)=0$ for $\vpar\geq0$ and $\vpar\leq0$, respectively.

\subsection{\gdb's two-fluid model}

We employed the drift-reduced two-fluid Braginskii model. These equations are obtained via a simplification of the collisionally-closed two-fluid equations assuming time variations that are slow compared to the ion gyro-frequency, $d/dt\ll\omegaci$. Under this assumption the lowest order perpendicular flows are the \ExB~($\vvExB=c\thinspace\bhat\times\grad{\phi}/B$) and diamagnetic flows ($\vvds=c\thinspace\bhat\times\grad{p_s}/(enB)$). The reduction also invokes quasineutrality, and discards the electron polarization drift and some terms smaller by a factor of $m_e/m_i$. We also do not include collisional drift terms~\cite{Madsen2016}. The result of such procedure~\cite{Francisquez2018a} is the following set of equations for the time rate of change of the plasma density $n$, generalized vorticity $\gvort$, mean field-aligned flows $\upars$ and temperatures $T_s$:
\begin{align}
\d{^en}{t}&=-\frac{2c}{B}\left[n\curv{\phi}-\frac{1}{e}\curv{p_e}\right]-n\delpar{\vpe}+S_n+\mathcal{D}_n \label{eq:nDot}
\end{align}
\begin{align}
\pd{\gvort}{t} &=-\div{\left(\frac{nc^2}{B^2\omegaci}\left[\phi,\v{h}\right]+\frac{nc\vpi}{B\omegaci}\delpar{\v{h}}\right)}+\frac{1}{e}\delpar{\cur} \nonumber \\
&\quad+\frac{2c}{eB}\left[\curv{p_e}+\curv{p_i}\right]-\frac{1}{3m_i\omegaci }\curv{G_i}+\mathcal{D}_\gvort \label{eq:wDot}
\end{align}
\begin{align}
\d{^e\vpe}{t}&=-\frac{1}{m_e}\left(\frac{\delpar{p_e}}{n}+0.71\delpar{T_e}-e\delpar{\phi}-e\etapar\cur\right) \nonumber \\
&\quad+\frac{2cT_e}{eB}\curv{\vpe}+\frac{2}{3}\frac{\delpar{G_e}}{m_en}-\frac{\vpe}{n}S_n+\mathcal{D}_{\vpe} \label{eq:vpeDot}
\end{align}
\begin{align}
\d{^i\vpi}{t}&=-\frac{1}{m_i}\left(\frac{\delpar{p_i}}{n}-0.71\delpar{T_e}+e\delpar{\phi}+e\etapar\cur\right) \nonumber \\ &\quad-\frac{2cT_i}{eB}\curv{\vpi}+\frac{2}{3}\frac{\delpar{G_i}}{m_in}-\frac{\vpi}{n}S_n+\mathcal{D}_{\vpi} \label{eq:vpiDot}
\end{align}
\begin{align}
\d{^eT_e}{t}&=\frac{2}{3}\frac{T_e}{n}\left[\d{^en}{t}+\frac{1}{T_e}\delpar{\kappapare\delpar{T_e}}+\frac{5n}{m_e\omegace}\curv{T_e}+\etapar\frac{\cur^2}{T_e} \right. \nonumber \\
&\left.\quad+\frac{0.71}{e}\left(\delpar{\cur}-\frac{\cur}{T_e}\delpar{T_e}\right)+\frac{1}{T_e}S_{E,e}-\frac{3}{2}S_n\right]+\mathcal{D}_{T_e} \label{eq:TeDot}
\end{align}
\begin{align}
\d{^iT_i}{t}&=\frac{2}{3}\frac{T_i}{n}\left[\d{^in}{t}+\frac{1}{T_i}\delpar{\kappapari\delpar{T_i}} \right. \nonumber \\
&\left.\quad-\frac{5n}{m_i\omegaci }\curv{T_i}+\frac{1}{T_i}S_{E,i}-\frac{3}{2}S_n\right]+\mathcal{D}_{T_i}. \label{eq:TiDot} 
\end{align}
Note that these fluid equations appear in Gaussian units, while the gyrokinetic model is written in SI units. Previous Braginskii simulations of Helimak, and some modern tokamak fluid simulations~\cite{Halpern2016}, do not evolve the ion temperature in cases where this is thought to be small but in this work we retain $T_i(\v{x},t)$. Equation~\ref{eq:wDot} evolves the generalized vorticity $\gvort=\div{nc\v{h}/(\omegaci B)}$, written in terms of the vector $\v{h}=\grad{\phi}+\left(\grad{p_i}\right)/(en)$. This equation has a contribution in terms of $G_s=\eta_0^s\left\lbrace2\delpar{\upars}+c\left[\curv{\phi}+\curv{p_s}/(q_sn)\right]/B\right\rbrace$, the gyroviscous part of the stress tensor with  $\eta_0^s$ being the viscosity of species $s$~\cite{Huba2013}. The time rate of change $d^s F/dt= \partial F/\partial t+c\left[\phi,F\right]/B+\upars\delpar{F}$ is given in terms of the Poisson bracket $[F,G]=\bhat\times\grad{F}\cdot\grad{G}$ and the parallel derivative $\delpar{F}=\bhat\cdot\grad{F}$. The effect of curvature of the magnetic field, $\kappacurv=-\hat{R}/R$, is captured by the operator $\curv{F}=-\bhat\times\kappacurv\cdot\grad{F}$. The coefficients $\kappa_{\parallel}^s$ and $\eta_{\parallel}$ are the parallel heat diffusivity and conductivity~\cite{Huba2013}. We use the notation $\cur=en(\vpi-\vpe)$ for the parallel current. 

Equations~\ref{eq:nDot}-\ref{eq:TiDot} include diffusion terms ($\mathcal{D}_F$) added for numerical stability consisting of both sixth-order perpendicular and second-order parallel diffusion. The latter is not always necessary for stability, but is needed in order to produce a physical $\kpar$ spectrum. There are also particle ($S_n$) and energy ($S_{E,s}$) sources (no momentum sources) given by
\begin{eqnal} \label{eq:gdbSrc}
S_n(x,z) &= \SOfl\exp\left[-\frac{\left(R-\xSrc\right)^2}{2\sigSrc^2}\right], \\
S_{E,e}(x,z) &= 1.87 \cdot \frac{3}{2}T_{e0} S_n(x,z), \\
S_{E,i}(x,z) &= 0.131\cdot \frac{3}{2}T_{i0} S_n(x,z),
\end{eqnal}
where $\SOfl=6.525\times10^{19}$. The form of these fluid source was chosen to follow the plotted \gkeyll~sources in figure~\ref{fig:Srcs}. Notice that their amplitudes are lower than those obtained from equations~\ref{eq:nSrc} and~\ref{eq:enerSrc} with $S_0=8.6\times10^{19}$ m$^{-3}$ s$^{-1}$; this is explained in section~\ref{sec:results}.

\begin{figure}
\includegraphics[width=0.48\textwidth]{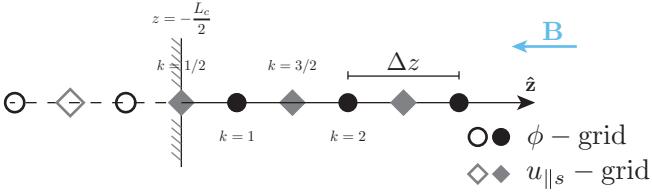}
\caption{\label{fig:GDBzGrid} Grids along the $z$ direction in \gdb~showing staggering of $\upars$ (diamonds) with the other quantities defined on the $\phi$-grid (circles). Here we only show a few cells near the upper sheath, at $z=-\Lc/2$. Cell centers are indexed by $k$.}
\end{figure}

The fluid equations are solved by the finite difference code \gdb, and the numerical details are described in previous publications~\cite{Zhu2018,Francisquez2019}. Here we only report on details of the numerical implementation relevant to the comparison with \gkeyll~and pertaining the Helimak geometry. The BCs in the radial direction are homogeneous Dirichlet for $\phi$ and $\gvort$, and even symmetry BCs for $n$, $T_{s}$ and $\upars$. These symmetric BCs are implemented by filling ghost cells so that there is symmetry about the wall surface (at $x=\pm\Lx/2$), but they do not enforce a zero-gradient at the first and last radial grid points since these are located half a grid spacing away from the wall. Hence, fluxes to the walls are allowed in \gdb. The fluid code uses a grid staggered in $z$ for the parallel velocities $\upars$ (see figure~\ref{fig:GDBzGrid}). Therefore the cell center coordinates of the $\upars$-grid are given by $z_{k}^{\upars}=(k-1)\Dz-\Lc/2$ for $k\in[1,\Nz+1]$, while the other quantities are defined on the $\phi$-grid with coordinates $z_k^\phi=(k-1/2)\Dz-\Lc/2$ with $k\in[1,\Nz]$. Since we impose the lower limit of the Bohm criterion as a sheath BC for $\upars$~\cite{Stangeby2000}, in practice this means that at the upper sheath, for example, we set (recall $\bhat=-\zhat$)
\begin{eqnal}
u_{\parallel i,k=1/2} &= -c_{s,k=1/2}=-\sqrt{\frac{T_{e,k=1/2}+T_{i,k=1/2}}{m_i}}, \\
u_{\parallel e,k=1/2} &= \begin{cases} -c_{s,k=1/2}\exp\left(\Lambda-\frac{e\phi}{T_e}\right)_{k=1/2} \quad \phi>0 \\
-c_{s,k=1/2}\exp\left(\Lambda\right) \qquad\qquad\quad\hspace{5pt} \phi\leq 0
\end{cases}
\end{eqnal}
where $\Lambda=\log\sqrt{m_i/[2\pi m_e(1+\tau)]}$, $\tau=T_{i0}/T_{e0}$ is the temperature ratio, and the temperatures and potential at $k=1/2$ are obtained via two-point linear extrapolation. We could instead impose the correct Bohm-sheath criterion $\vpi\leq-c_s$ at this sheath by using homogeneous Neumann BCs whenever the local flow is supersonic, but the intention here is to employ techniques used by Braginskii solvers in the past. There are also more sophisticated sheath BCs accounting for oblique incidence of the magnetic field~\cite{Loizu2012}, which will be interesting to consider in future fluid-gyrokinetic comparisons.

Evolving the values of $n$, $\phi$ and $T_s$ in the first and last cells along $z$ also requires parallel BCs for these quantities (and $\gvort$). While many choices exist, researchers often choose those which exhibit best numerical stability. Given that we know the direction of the flow at the sheath entrance, we fill the $z$-ghost cells (empty circles and diamonds in figure~\ref{fig:GDBzGrid}) such that an upwind stencil ensues. For example $T_{i,k=0}=3(T_{i,1}-T_{i,2})+T_{i,3}$. The calculation of parabolic terms ($\propto \delpar{^2}$) are computed using homogeneous Neumann BCs, except for the heat diffusivity terms discussed below.

We will distinguish between three different boundary conditions for the heat diffusivity terms ($\propto \kappa_{\parallel}^s$) in the temperature equations. Our first choice will be to use homogeneous Neumann BCs, which lead to a zero-heat flux condition ($q_{\parallel s}=0$) at the sheath. This is a common choice as it provides superior numerical stability. However, finite conductive heat-fluxes entering the sheath are measured experimentally, so we will also explore the effect of $q_{\parallel s}\neq0$ BCs. We implement the latter by imposing
\begin{eqnal} \label{eq:heatFlux}
q_{\parallel s}=-\kappa_\parallel^s\delpar{T_s}=\pm \gamma_s n\upars T_s,
\end{eqnal}
where the upper (lower) sign corresponds to the top (bottom) sheath, and $\gamma_s$ is the sheath transmission coefficient. The heat transmission coefficients employed here will be
$\gammae=2+|e\phi|/T_e$ and
$\gammai=2.5T_i/T_e$. Note that this expression neglects the convective and frictional parts of the heat-flux. We will thus consider a third finite heat-flux BC which does account for them in the electron channel ($q_{\parallel e}^{\mathrm{tot}}\neq0$), consisting of
\begin{eqnal} \label{eq:qeTot}
q_{\parallel e}^{\mathrm{tot}} &= n_e\upare\left(\frac{5}{2}T_e+\frac{1}{2}m_e\upare^2\right) - \kappapare\delpar{T_e} \\
&\quad+0.71n_eT_e\left(\upare-\upari\right) = \pm\gamma_en_e\upare T_e,
\end{eqnal}
although in \gdb~the $m_e\upare^2/2$ term was neglected as small compared to $5T_e/2$. In practice the finite heat-flux BCs are applied by filling the ghost cells accordingly; for example, at the upper sheath the BC in equation~\ref{eq:heatFlux} entails
\begin{eqnal} \label{eq:qFluxBC}
\left(\ln T_s\right)_{k=0} &= \left(\ln T_s\right)_{k=1}-\frac{\Dz}{\kappa_{\parallel}^s}\gamma_s n \upars\Bigg|_{k=1/2}.
\end{eqnal}

Note that \gdb~and many other Braginskii codes evolve the logarithms of the density and the temperatures. This is a widely used technique to guarantee the positivity of such quantities. However a pitfall of this approach is that it becomes more challenging to have a conservative scheme. It is common to ameliorate the effects of non-conservation, turbulent cascades and the lack of upwinding with the use of additional numerical diffusion. Some Braginskii codes write such diffusive terms in conservative form~\cite{Stegmeir2019}, but that requires either an explicit treatment or solution to a nonlinear elliptic problem.
In \gdb~we use diffusive terms of the form~\cite{Smith1997,Francisquez2019}
\begin{equation}
\mathcal{D}_F=\chi_x\pd{^6F}{x^6}+\chi_y\pd{^6F}{y^6}+\chi_\parallel\nabla_\parallel^2F
\end{equation}
discretized with second-order centered finite differences. In order to treat the perpendicular diffusion implicitly we apply it on the logarithm of $n$ and $T_s$.

The code solves a normalized form of equations~\ref{eq:nDot}-\ref{eq:TiDot}, given in appendix~\ref{sec:gdbNorm}. We make several additional approximations in order to make a comparison with a fluid model representative of those used by other Braginskii solvers~\cite{Zhu2018}. The first of these is the Boussinesq approximation, which \gdb~is usually run without: $\gvort=\div{n_{e0}c\v{h}/(\Omega_{i0} \BO)}$. As noted in this definition of the vorticity, we will examine \gdb~simulations without the variation in the magnetic field amplitude, $B=B_0$, and will likewise use $R=R_0$. Often this approximation is made in the simulation of tokamak annuli because the radial extent is small, and the impact of the radial variation in $B$ is thought to be small. We also disregard the spatial variation of the $\etapar$, $\eta^s_0$ and $\kappa_\parallel^s$, though \gkeyll~retained the spatial dependence in the collisionality: $\nu_{sr}=\nu_{sr}(x)$. These changes can modify the simulation significantly depending on which diagnostic one looks at~\cite{Francisquez2018a}, but the intention here is to use assumptions and simplifications  typical in Braginskii simulations found in the literature. Reporting on the effects of additional levels of complexity is left for future publication.

\section{Simulation results} \label{sec:results}

\begin{figure}
\includegraphics[width=0.48\textwidth]{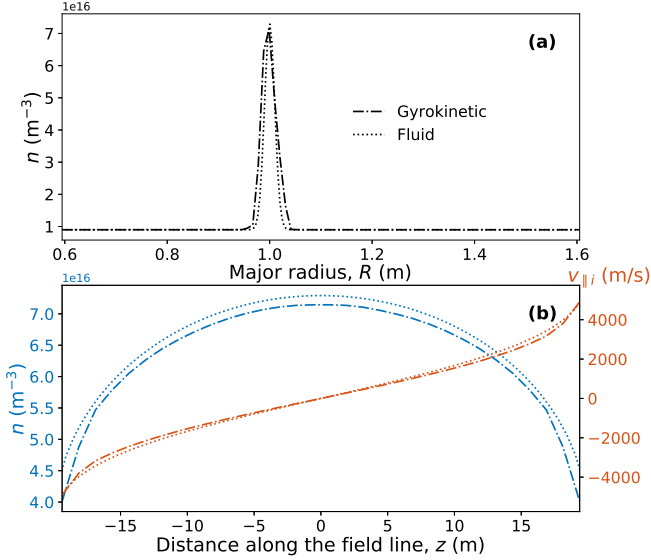}
\caption{\label{fig:ICs} Density and ion parallel velocity initial conditions in the gyrokinetic and fluid simulations. (a) Variation along $R$, and (b) field-aligned variation.}
\end{figure}

In this section we describe the data from the fluid and gyrokinetic simulations of the Texas Helimak, and in the following section offer additional analysis and discussion. \gkeyll~simulations used shifted-Maxwellian initial conditions (ICs) that resemble the expected steady state profiles, except for the radial density profile which followed the Gaussian source~\cite{Bernard2019,Shi2019}. \gdb~was provided with similar ICs. The initial density of \gkeyll~was matched in \gdb~with the following IC:
\begin{eqnal}
n(x,z,t=0) &= \frac{n_p}{4.9778}\left\lbrace1.0661\exp\left[-\frac{\left(R-\xSrc\right)^2}{2\sigSrc^2}\right]+0.1\right\rbrace \\
&\qquad\times\frac{1+\sqrt{1-z^2/(\Lc/2)^2}}{2}.
\end{eqnal}
A small density floor was added to avoid positivity issues in \gkeyll~at early times. This initial profile was perturbed randomly with small amplitude fluctuations. The radial and field-aligned variation of this density profile is shown in figure~\ref{fig:ICs}. Also shown there is the initial parallel ion velocity, given approximately as
\begin{equation}
\vpi(z,t=0) = 1.275\thinspace\cseO\thinspace \frac{1-\sqrt{1-z^2/(\Lc/2)^2}}{z/(\Lc/2)}.
\end{equation}
The electron parallel velocity was essentially zero at $t=0$, and there was a small temperature gradient which in \gdb~we modeled as
\begin{eqnal}
T_e(x,t=0) &= \frac{0.96}{1+0.22\,x/R_0}\thinspace T_{e0}, \\
T_i(x,t=0) &= \frac{1.17}{1+0.07\,x/R_0}\thinspace T_{i0}.
\end{eqnal}
In \gdb~the initial vorticity was set to zero, while \gkeyll~computes the initial electrostatic potential from solving the Poisson equation~\ref{eq:poisson}.

These initial conditions were discretized, and the ensuing solutions obtained, on a \gkeyll~phase-space mesh of $48\times24\times16\times10\times5$ cells and a piecewise-linear basis ($p=1$), while the \gdb~configuration-space grid employed $256\times128\times64$ points. The velocity space domain was given by $[-\txtSb{v}{s}{,max},\txtSb{v}{s}{,max}]\times[0,3m_s\txtSb{v}{s}{,max}^2/(16\BO)]$, where $\txtSb{v}{e}{,max}=4v_{te0}=4\sqrt{T_{e0}/m_e}$ and $\txtSb{v}{i}{,max}=6\cseO=6\sqrt{T_{e0}/m_i}$. The resolution of the kinetic simulation may seem coarse, but the radial spectrum of the turbulence is well converged~\cite{Bernard2019}. The gyrokinetic simulation used $180\thinspace000$ CPU-hours on Skylake nodes of the Texas Advanced Computing Center's Stampede2 cluster to reach 16 ms, while a fluid calculation required $16\thinspace600$ CPU-hours on MIT's Engaging cluster's Intel Xeon E5 2.1 GHz nodes (although the heat-flux BC choice can increase runtime by a factor of two). The cost in units of CPU-hours per milisecond, per degree of freedom was only 16\% higher for \gkeyll.

\begin{figure}
\includegraphics[width=0.48\textwidth]{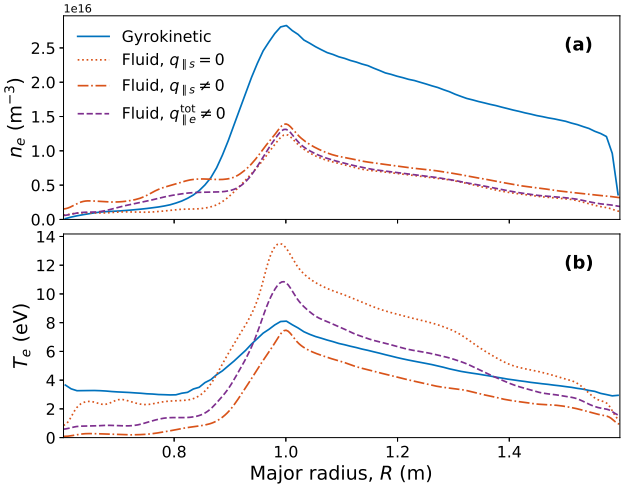}
\caption{\label{fig:profiles} Mean electron density (a) and temperature (b) profiles at $z=-\Lz/2$ as a function of major radius in fluid and gyrokinetic simulations. Fluid profiles are given for the simulations with boundary conditions: $\delpar{T_s}=0$ (effectively $q_{\parallel s}=0$), $q_{\parallel s}\neq0$ and $q_{\parallel e}^{\mathrm{tot}}\neq0$.}
\end{figure}

\subsection{Profiles, heat-flux BCs and sourcing}

These simulations begin with a period of linear growth in which fluctuation amplitudes increase due to free energy from the pressure gradient and the curvature of the magnetic field. This is typical of curvature-driven modes such as interchange or ballooning modes. Fluctuations are visually imperceptible in the short linear phase, but as they grow, they self-organize into radial streamers and mushroom or blob-like structures visible after $\sim150~\mu$s. As these formations saturate, they modify the perpendicular fluxes and alter the parallel transport. Eventually the simulation reaches a quasi-steady state in which the sources are balanced by the transport and the sheath losses and fluctuation levels saturate. Once in quasi-steady state we begin analyzing plasma profiles and turbulence properties. For example, by $y$- and time-averaging electron density and temperature at $z=-\Lz/2$ we obtain the profiles in figure~\ref{fig:profiles}. Time averages were carried out in the $10-16$ ms window, which is larger than the $\Lc/(2\cseO)\sim4$ ms ion transit time and at which point turbulence level is saturated. The profiles presented here, unless stated otherwise, are measured at the bottom sheath ($z=-\Lz/2$) because that is one place where probe measurements are taken in the experiment. 

Figure~\ref{fig:profiles} shows that the peak density is roughly twice as large in the \gkeyll~calculation than in the \gdb~simulation with $q_{\parallel s}=0$ BCs. There is also a drop in \gkeyll's electron density near the radial boundary which is not present in \gdb. As a result of the gyrokinetic simulation approaching sheath ambipolarity the potential tends to $\phi\sim\Lambda T_e$ in the interior of the domain but $\phi=0$ at the boundary, so electrons near the radial boundaries are lost to the end plates very quickly because there is little sheath potential to confine them. As mentioned in section~\ref{sec:models}, there are differences in the radial BCs (zero radial flux for the gyrokinetic code vs. finite radial flux for the fluid code) so there is more radial flux to offset rapid parallel losses in the fluid code than in the gyrokinetic code.

Besides differences in density profiles, figure~\ref{fig:profiles} also reveals a \gdb~$T_e$ that is 67\% greater than the gyrokinetic $T_e$. The \gkeyll~sheath BCs allow a particle flux out of the simulation that carries heat with it, while our first \gdb~simulation explicitly imposed $\delpar{T_s}=0$ in the heat-flux terms of temperature equations~\ref{eq:TeDot}-\ref{eq:TiDot}. In order to allow for a finite heat-flux into the sheath we implemented the BCs in equation~\ref{eq:heatFlux} in \gdb. This lowered the electron temperature at the sheath, and coincidentally nearly matched \gkeyll's peak $T_e$ (see the orange dash-dot line in figure~\ref{fig:profiles}b).

The considerable difference in the electron temperature of the $q_{\parallel s}=0$ and $q_{\parallel s}\neq0$ fluid simulations is a result of the BCs in equation~\ref{eq:heatFlux} extracting a disproportionate amount of heat. Such BCs were setting the entire electron heat flux ($\gamma_en_e\upare T_e$) to equal the conductive component, while a more appropriate BC takes into account the convective and frictional components as well. Such is the case of the $q_{\parallel e}^{\mathrm{tot}}\neq0$ BCs in equation~\ref{eq:qeTot}, which effectively sets the conductive heat flux to be smaller than in the $q_{\parallel s}\neq0$ simulation. The result is a slower release of heat through the sheath and thus a larger electron temperature across the plasma compared to $q_{\parallel s}\neq0$ BCs (see the purple dashed line in figure~\ref{fig:profiles}). The gamut of zero and finite heat-flux BCs has been employed by the body of Braginskii codes in the past. Such codes sometimes set $q_{\parallel s}=0$~\cite{Halpern2016}, but other times they use variations of finite heat-flux BCs. For example, the $q_{\parallel s}\neq0$ BCs  in equations~\ref{eq:heatFlux}-\ref{eq:qFluxBC} are sometimes used in other Braginskii codes, albeit with a sheath transmission coefficient ($\gamma_e$) that takes into account the $(5/2)T_e$ term in equation~\ref{eq:qeTot}~\cite{Xia2013,Stegmeir2019}. Although $q_{\parallel e}^{\mathrm{tot}}\neq0$ BCs are more accurate, in what follows we only use the $q_{\parallel s}\neq0$ BCs of equation~\ref{eq:heatFlux} because on this occasion, in combination with the other approximations made, they produced a peak fluid $T_e$ closer to the maximum gyrokinetic $T_e$. 
\begin{figure}
\includegraphics[width=0.49\textwidth]{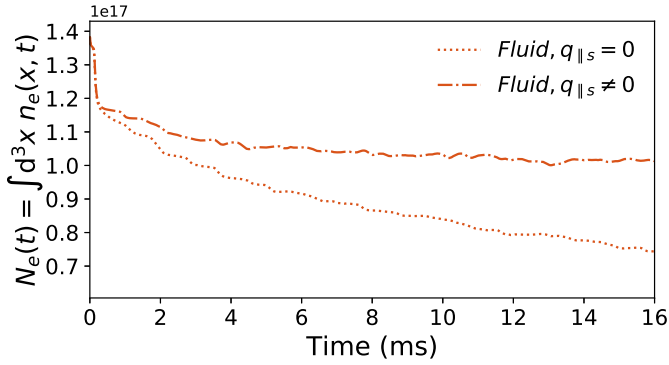}
\caption{\label{fig:NvT} Time trace of the total number of electrons in the fluid simulations with zero heat-flux (dotted line) and finite heat-flux (dash-dot line) BCs.}
\end{figure}

Another observation on the consequences of $q_{\parallel s}$ BCs is that despite the temperature drop caused by the $q_{\parallel s}\neq0$ heat-sink, the peak density remained unaltered (compare orange dotted and dash-dot lines in figure~\ref{fig:profiles}a). Since sheath physics play an important role in this system, one might have expected that lowering $T_e$ would cause $c_s$ to decrease and, hence, the outflow of particles to slow down and thus the density to increase. The change between the two orange lines in figure~\ref{fig:profiles}a does indicate an increase in the average density, and the particle loss rate did decrease when the heat-flux was allowed to be finite at the sheath entrance. This change is demonstrated in figure~\ref{fig:NvT}, showing the time trace of the number of electrons throughout the fluid simulations. At the end of the 16 ms period the $q_{\parallel s}\neq0$ simulation has nearly ${\sim}34\%$ more electrons in it than the zero heat-flux counterpart. The fact that the peak density remained constant suggests that the perpendicular transport and  conservation errors jointly increased to meet the weakening parallel losses.

\begin{figure}
\includegraphics[width=0.49\textwidth]{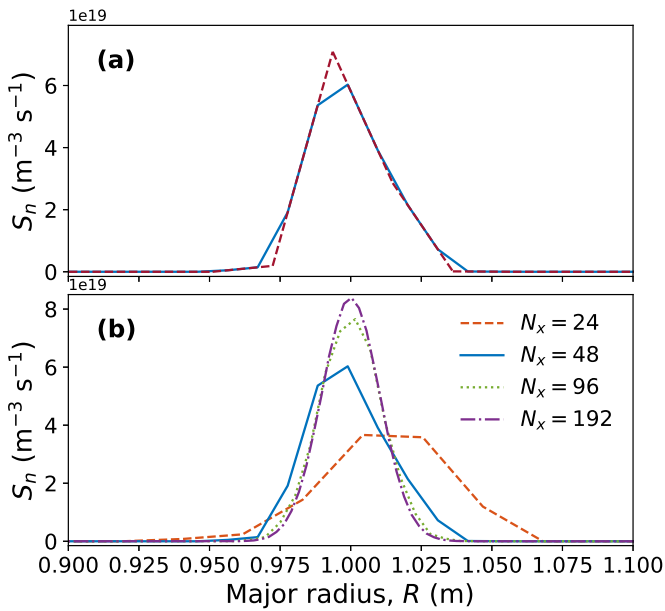}
\caption{\label{fig:srcDG}(a) \gkeyll~particle source ($S_n$) plotted in~figure~\ref{eq:src}a using cell-center values on a grid with $\Nx(p+1)$ (solid blue), and its discontinuous piecewise-linear representation (dashed red). (b) \gkeyll~$S_n$ interpolated onto an $\Nx(p+1)$ grid with varying resolution.}
\end{figure}

It is still surprising that the gyrokinetic simulation yielded a much higher peak density and an entirely different density profile. A more careful examination of the sources reveals that the \gdb~simulation actually used a smaller particle fuel rate than that in \gkeyll. Part of the reason for this relates to the nature of the DG representation. The source in \gkeyll~is a DG representation of equation~\ref{eq:src} constructed by evaluating such function at cell-boundary nodes. The particle source in \gdb~(equation~\ref{eq:gdbSrc}) was chosen to match the plotted \gkeyll~source (in figure~\ref{fig:Srcs}). The \gkeyll~lines in these plots were created by evaluating the cell-center value on a grid with $\Nx(p+1)$ cells. But note that the maximum value does not necessarily occur at these plotted coordinates, or at the cell nodes where the function was evaluated to construct the DG representation. We can plot the local piecewise linear representation (dashed red line in figure~\ref{fig:srcDG}a) to confirm that \gkeyll~actually has a higher value source than we had previously interpreted. Plotting and post-processing DG data can sometimes require subtle consideration of the underlying higher-order nature of the solution in order to avoid these errors. Another way to appreciate this nuance is by discretizing and plotting $S_n$ with increasing resolution (figure~\ref{fig:srcDG}b): the amplitude converges towards the $S_0=8.6\times10^{19}$ m$^{-3}$ s$^{-1}$ mentioned in section~\ref{sec:models}. The dashed orange line is significantly lower amplitude because at this coarse resolution the maximum of the Gaussian source lies farther from and drops off fast towards the cell-boundary nodes where equation~\ref{eq:src} was evaluated to construct the DG representation. The plots with $\Nx=48$ do not imply that the \gkeyll~simulation used a lower source than it should have been, but rather highlight that the projection of a function onto the DG basis needs to be carefully analyzed.

\begin{figure}
\includegraphics[width=0.49\textwidth]{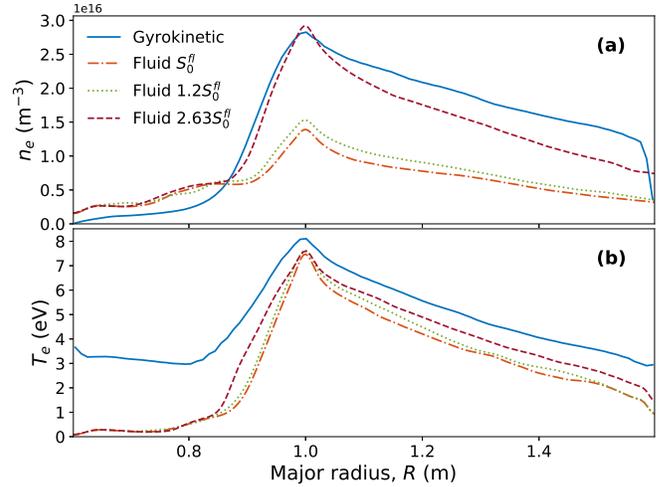}
\caption{\label{fig:profilesVsrc}Electron density (a) and temperature (b) in the gyrokinetic simulation (solid blue) compared to those in $q_{\parallel s}\neq0$ fluid simulations with the sources increased by 20\% and 163\% from the initial $\SOfl$.}
\end{figure}

Rather than roughly matching the plotted fluid and gyrokinetic sources, it is more suitable to guarantee that volume integrals of such sources agree. In order to match the volume integral of \gkeyll's density source rate (using only its cell averages and a composite trapezoidal method) it became necessary to increase the particle source rate in the fluid calculation (equation~\ref{eq:gdbSrc}) by 20\%, and the heat sources by 22.94\%. The fluid simulation was re-run with this $1.2\SOfl$ source amplitude, keeping other parameters fixed. In figure~\ref{fig:profilesVsrc} we show the effect on the time- and $y$-averaged radial density and electron temperature profiles and compare the \gdb~result (using $q_{\parallel s}\neq0$ BCs) with \gkeyll's. Increasing the sources by 20\% did not alter the electron temperature profile significantly; only slight modifications are seen across the entire radius. The peak density only increased by about 16\%, and still remained significantly far away from the gyrokinetic density profile. The fact that the cross-field turbulent spreading does not appear to increase, because the boundary values and profile were relatively unchanged, suggests that the parallel transport is strong and likely convects any excess particle input out the sheaths. Increasing the sources in \gdb~by 50\% and by 125\% failed to match the \gkeyll~profiles. In order to approach the gyrokinetic peak density we had to augment the fluid sources by a factor of $\sim2.63$ (163\%), shown by the red dashed line in figure~\ref{fig:profilesVsrc}. It is possible to slightly adjust the \gdb~particle and temperature sources independently in order to match the gyrokinetic peak values of both $n$ and $T_e$.

With the $2.63\SOfl$ source, the difference in profile maxima for the fluid and gyrokinetic simulations is 0.6\% for $n$ and 8.6\% for $T_e$, and substantial differences can be seen in their profiles. Perpendicular particle transport more effectively widens the density profile in \gkeyll, resulting in higher densities and lower gradients on the low-field side. There is also a sharp drop in the gyrokinetic electron density at $R\approx1.6$ m that is absent in the \gdb~data, which might be caused the differences in the radial BCs; \gdb~is allowing radial fluxes to the wall, while \gkeyll~is not (section~\ref{sec:models}). On the outboard side both codes produce a similar $T_e$ profile, albeit shifted down by almost 1 eV in the fluid simulation. On the high-field side, the $T_e$ profile is more than ten times larger in the \gkeyll~data. There is a minimum of $T_{e,\text{min}}=1.7$ eV that can be resolved by \gkeyll~in order to maintain a positive distribution function with this resolution~\cite{Shi2017}, but even experimental data suggests $T_e\sim 2.5$ eV near $R\sim 0.8$ m (see figure 8b in~\cite{Bernard2019}). The extremely low \gdb~high-field side $T_e$ is caused by the choice of $q_{\parallel s}\neq0$ BCs, which as explained earlier, can cause the electrons to cool too rapidly because they neglect effects from convective and friction terms (compare dashed purple and dash-dot orange lines in figure~\ref{fig:profiles}). Accounting for these terms in finite heat-flux BCs, as well as using a spatially varying heat conductivity ($\kappapare$) will substantially increase the electron temperature on the high-field side.

\begin{figure}
\includegraphics[width=0.49\textwidth]{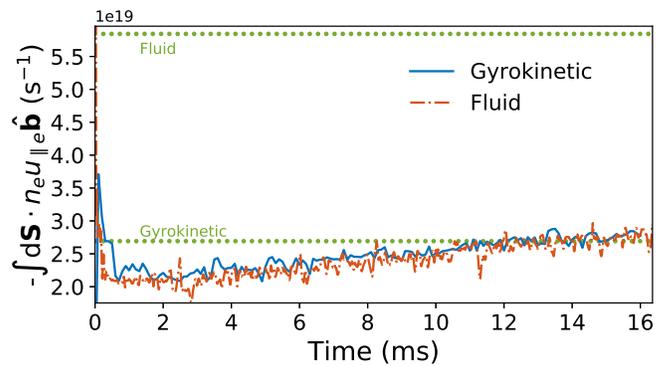}
\caption{\label{fig:GammaNtt}Time trace of the parallel electron particle flux integrated over the sheath surface in the gyrokinetic (solid blue) and $2.63\SOfl$ fluid (dash-dot orange) simulations. The volume integrals of their particle source is given by the green circles.}
\end{figure}

In addition to the differences between the gyrokinetic and the $2.63\SOfl$ fluid simulations, we note that both fluid and gyrokinetic global profiles (figure~\ref{fig:profilesVsrc}) exhibit discrepancies compared to the experimentally measured, bottom-sheath equilibrium profiles~\cite{Bernard2019}. Although the peak experimental density is close in magnitude to the simulated results, the experimental radial profiles have stronger local gradients, indicating weaker cross-field transport. Also, the experimental $T_e$ profile has a higher peak value located at larger radii.


\subsection{Particle conservation in \gdb}

The partial agreement we did attain in fluid and gyrokinetic density profiles came at the expense of fueling \gdb~2.1875 times more strongly ($2.63\SOfl$), compared to the simulation that matched the volume integrated sources ($1.2\SOfl$). One may first suspect that the gyrokinetic model of conducting sheath BCs yields a slower outflow, but that is not the case as the integral of $n\v{\upare}$ over the sheaths demonstrate (figure~\ref{fig:GammaNtt}). In addition to the radial particle fluxes to the wall allowed in \gdb, another candidate explanation for the additional particle loss is that \gdb~is not conservative. Formulation errors (e.g. approximations to $B(R)$ and geometric factors), discretization errors (e.g. from non-conservative finite differences using $\ln n_e$ instead of conservative finite differences using $n_e$), and numerical diffusion can conspire to break particle conservation. Figure~\ref{fig:GammaNtt} is evidence that non-conservation errors can be $\mathcal{O}(1)$; even though the volume integrated density is in quasi-steady state the parallel flux to the end plates is only about half of the input source in the fluid code, meaning that the other half of the particles are being lost mostly due to some other errors, either due to the formulation of the fluid equations or numerical errors.

\begin{figure}
    \centering
    \includegraphics[width=0.49\textwidth]{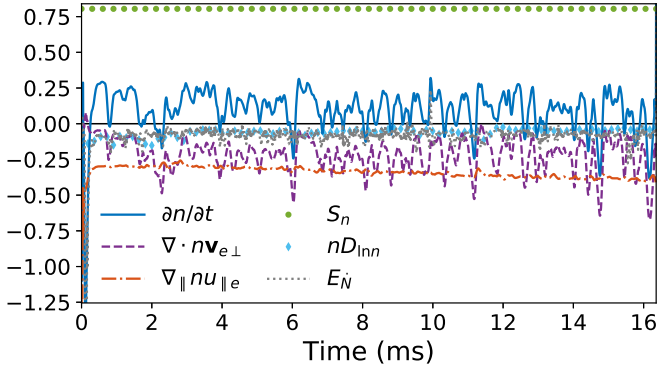}
    \caption{Volume integrated terms in the normalized density equation~\ref{eq:nDot} in the \gdb~simulation with $q_{\parallel s}\neq0$ and $2.63\SOfl$ (see normalization in section~\ref{sec:gdbNorm}). The blue line ($\partial n/\partial t$) is computed using a centered finite difference in time between snapshots 6.06 $\mu$s apart, and the grey dotted line shows $\partial n/\partial t$ minus all the other lines. A 96 $\mu$s moving average was applied to all lines before plotting.}
    \label{fig:ldenBalance}
\end{figure}

The aforementioned diffusion terms are nonetheless thought to be small in these \gdb~runs (see reported values in appendix~\ref{sec:gdbNorm}), but its true impact depends on the resolution and the turbulent scales generated by the regime one is simulating. We can examine the volume integral of each of the terms in the density equation (figure~\ref{fig:ldenBalance}) and
confirm the relative smallness of the diffusion terms (blue diamonds, although in this diagnostic the contribution from perpendicular diffusion was computed using the density outputted every 400 time steps and is likely to differ some from the impact of the true, implicit perpendicular diffusion). This analysis also confirms a concerning imbalance between parallel losses and sources: the former only accounts for 54\% of the latter towards the end of the simulation (compare orange dash-dot line and green circles). This gap is all the more puzzling since the particle accounting error
\begin{eqnal}
E_{\dot{N}} = \avg{\pd{n}{t}\Big|_{FDT}}_{xyz} &- \left(- \avg{\div{n\v{v}_{e\perp}}}_{xyz} - \avg{\delpar{n\upare}}_{xyz} \right. \\
&\qquad\left.+ \avg{S_n}_{xyz} + \avg{n\mathcal{D}_{\ln n}}_{xyz}\right),
\end{eqnal}
with the first term measured with finite differences in time and $\div{n\v{v}_{e\perp}}=(c/B)[\phi,n]+(2c/B)\left[n\curv{\phi}-\curv{p_e}/e\right]$, appears to be negligible (dotted grey line). This seemingly small error is only possible because $\partial n/\partial t$ is on average not zero, and because the volume integral of $\div{n\v{v}_{e\perp}}$ does not vanish despite the BCs on $\phi$ (homogeneous Dirichlet in $x$ and periodic in $y$).

A non-vanishing $\int\text{d}^3\v{x}\,\div{n\vvExB}$ may indirectly affect the parallel sheath losses, so we are interested in ensuring this basic feature. We illustrate this with a concrete simpler example, neglecting the vertical field in the Helimak and considering a purely toroidal field with $B$ in the $-\varphi$ direction over a small, periodic extent in $y$ confined by perfectly conducting walls in $x$. Adopt a coordinate system $(x,y,z)$ related to cylindrical coordinates by $(x,y,z) = (R,Z,-R_0\varphi)$, where $Z$ is the vertical coordinate and $\varphi$ the toroidal angle. In this case the volume element is then $\text{d}R\,\text{d}Z\,R\, \text{d}\varphi$ and we write this as
$\text{d}R\,\text{d}Z\,R_0\, \text{d}\varphi(R/R_0)=\text{d}x\,\text{d}y\,\text{d}z(R/R_0)=\text{d}^3\v{x}(R/R_0)$. The \ExB~particle balance entails
\begin{eqnal} \label{eq:nBalanceExB}
\pd{N_e}{t} &= - \int\text{d}^3\v{x} \,\frac{R}{R_0}\left\lbrace\frac{c}{B}\left[\phi,n_e\right]+\frac{2c}{B}n_e\curv{\phi}\right\rbrace, \\
&= - \int\text{d}^3\v{x} \,\frac{R}{R_0}\frac{c}{B}\left\lbrace\left(\pd{\phi}{y}\pd{n_e}{x}-\pd{\phi}{x}\pd{n_e}{y}\right)+2n_e\curv{\phi}\right\rbrace, \\
&= - \int\text{d}^3\v{x} \,\frac{R}{R_0}\frac{c}{B}\left\lbrace\left[\pd{}{x}\left(\pd{\phi}{y}n_e\right)-\pd{}{y}\left(\pd{\phi}{x}n_e\right)\right] \right.\\
&\left.\qquad\qquad\qquad\qquad+\frac{2}{R}n_e\pd{\phi}{y}\right\rbrace, \\
&= - \int\text{d}x\,\text{d}y\,\text{d}z \,\frac{R}{R_0}\left[\frac{c}{B}\pd{}{x}\left(\pd{\phi}{y}n_e\right)+\frac{2c}{BR}n_e\pd{\phi}{y}\right],
\end{eqnal}
where we made use of the periodic BCs along $y$ and the equality of mixed partials. For particle number to be conserved in this isolated \ExB~system it must be that the first term in equation~\ref{eq:nBalanceExB} cancels the second. 

\begin{figure}
    \centering
    \includegraphics[width=0.49\textwidth]{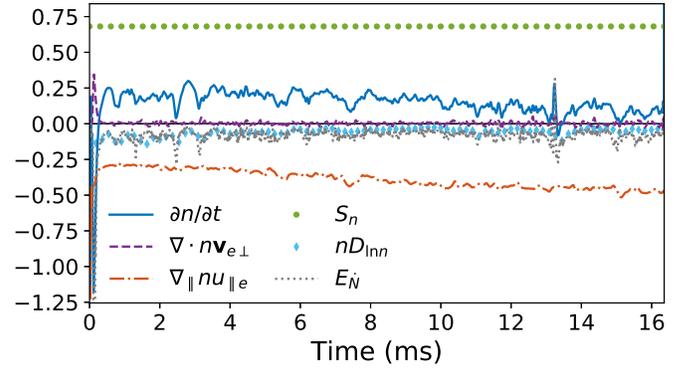}
    \caption{Similar to figure~\ref{fig:ldenBalance}, this plot shows volume integrated terms in the normalized density equation~\ref{eq:nDot}, but for a \gdb~simulation with $q_{\parallel s}\neq0$, $2.45\SOfl$ and keeping $R=R(x)$ and $B=B(R)$ as the analysis in equation~\ref{eq:nBalanceExB} suggests. We can see that by including these dependencies the volume-integrated perpendicular flux does vanish. A lower source amplitude ($2.45\SOfl$) was chosen because it was sufficient to produce a comparable \gdb~simulation with a plasma in a similar parameter regime as the \gkeyll~simulation.}
    \label{fig:ldenBalanceR}
\end{figure}

In the Helimak scenario such cancellation leads to the balance between sources and parallel losses, but this balancing was only partial in \gdb~because of modifications to the treatment and analysis of the \ExB~terms compared to what is in equation~\ref{eq:nBalanceExB}. The equations used in \gdb~so far account for the variation in $B$ in the curvature term ($1/(BR)=1$ in the second term of equation~\ref{eq:nBalanceExB}), but they replace $B\to\BO$ in the \ExB~nonlinearities ($c/B\to c/\BO$ in the first term of equation~\ref{eq:nBalanceExB}). This approximation was invoked in some annulus studies of tokamak SOLs because the variation in $R$ across that domain was thought to be small~\cite{Halpern2014,Francisquez2017}. It is often thought to be a small correction on the turbulent time-scale $\sqrt{L_{pe} R/2}/\cseO\simeq\sqrt{0.27\,\text{m}\cdot1.2\,\text{m}/2}/(3.5\times10^3\,\text{m/s}) = 0.115$ ms, but these errors build up and become significant on the confinement time scale $\Lc/\cs\approx 8$ ms. Furthermore, as Braginskii codes increasingly model larger radial domains the thin annulus approximation incurs ever larger inaccuracies. For the Helimak case presented here $B$ changes by a factor of $8/3\approx2.7$, and setting $1/B=1$ in the \ExB~term and $R/R_0=1$ in the equivalent particle balance shown in figure~\ref{fig:ldenBalance} causes the volume integrals of $n\div{\vvExB}$ and $\vvExB\cdot\grad{n}$ to not cancel each other, and $\int\text{d}^3\v{x}\,\div{n\v{v}_{e\perp}}$ to not vanish. We can estimate the relative size of this error contribution, defining $\avg{f}=\int\text{d}^3\v{x}\,f/(\int\text{d}^3\v{x}$), as
\begin{eqnal}
\frac{\int\text{d}^3\v{x}\,\frac{2c}{BR}n_{e1}\pd{\phi}{y}}{\int\text{d}^3\v{x}\,\delpar{n_e\upare}} &\sim \frac{\Lx\Ly\Lc\avg{\frac{2c}{BR}n_{e1}\pd{\phi_1}{y}}}{2\Lx\Ly n_{e0}\cseO} \sim \frac{\rhos}{R}\frac{\Lc}{L_{\perp}}\avg{\frac{n_{e1}}{n_{e0}}\frac{e\phi_1}{T_{e0}}}, 
\end{eqnal}
which can result in $\mathcal{O}(1)$ accumulated errors (we assumed maximally out of phase density and potential fluctuations of the same magnitude and estimated $\partialy\phi_1\sim L_{\perp}^{-1}\phi_1$, with $L_{\perp}$ a characteristic fluctuation perpendicular length scale). Note that even in a tokamak SOL, where $\rhos/R$ is small, this error can be significant because the parallel connection length can be very large and $\mathcal{O}(1)$ perturbations can occur. Had we also included the $1/B_{(R)}$ factors in the \ExB~nonlinearities we would find that the diffusion terms and the errors in particle balance (equivalent of blue diamonds and dashed grey line in figure~\ref{fig:ldenBalance}) account for about 21\% of the plasma injected, the non-zero $\partial n/\partial t$ for $\sim9\%$ and the rest is lost to the sheaths (see figure~\ref{fig:ldenBalanceR}). For context note that the latest version of most Braginskii fluid codes have the capability to keep the full Jacobian factors that improve conservation properties, and in recent years there has been more attention paid to implementing fluid algorithms with good conservation properties~\cite{Halpern2016,Tamain2016,Dudson2017,Stegmeir2018}.

\begin{figure*}
\includegraphics[width=\textwidth]{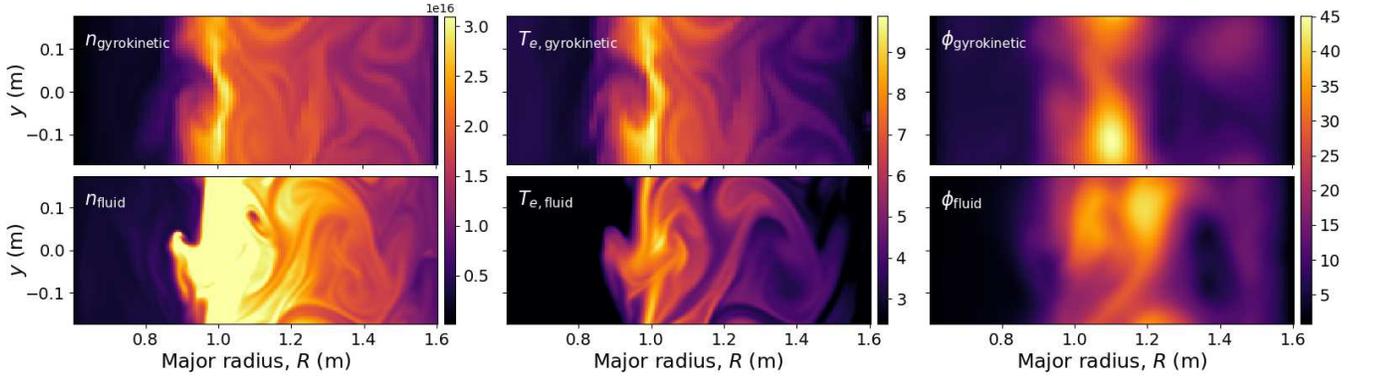}
\caption{\label{fig:snapshots}Snapshots of the $z=0$ plasma density (left column, in m$^{-3}$), electron temperature (center column, in eV) and electrostatic potential (right column, in volts) in the gyrokinetic (top row) and fluid (bottom row) simulations at $t=10$ ms. Colors scaled by gyrokinetic data.}
\end{figure*}

\subsection{Comparison of flows and turbulence in \gdb~and \gkeyll}

The $R\to R_0$ approximation made in these \gdb~simulations (except for that in figure~\ref{fig:ldenBalanceR}) had been used in some previous Braginskii turbulence work~\cite{Halpern2014,Francisquez2017}, and that with an enhanced source of $2.63\SOfl$ produced approximately the same regime as the gyrokinetic simulation. Hence, in what follows we compare \gkeyll~results with such \gdb~simulation ($2.63\SOfl$, $q_{\parallel s}\neq0$ and $1/B\to1$ in \ExB~convective terms) in more detail. A snapshot of the plasma density, electron temperature and electrostatic potential at 10 ms is given in figure~\ref{fig:snapshots} (colors set by the extrema in the gyrokinetic data). Instantaneous maxima can be very different in the two simulations even if the $y$ and time averaged profiles have similar maximum values, hence the bright yellow region in the snapshot of the fluid density. This maximal region suggests that radial transport is weaker there, while the \gkeyll~simulation appears to spread out the plasma radially more effectively. This is consistent with the average $n$ profile in figure~\ref{fig:profilesVsrc}. Furthermore, fluctuations in \gdb~seem have finer-scale structure, suggestive of a different $k_\perp$ spectrum and perhaps a smaller correlation length. The difference in radial turbulent spreading is also visible in the $T_e$ snapshot of figure~\ref{fig:snapshots}. The right-most column of this figure depicts relatively smooth $\phi$ profiles that do not resemble the plasma density fluctuations, indicating a significant departure from adiabaticity.

Looking at these snapshots a reader may be inclined to think that the gyrokinetic simulation is not well-resolved or that it is too diffusive. The smoother, larger perpendicular scales of the fluctuations in the gyrokinetic pictures of figure~\ref{fig:snapshots} are likely not a product of numerical diffusion since \gkeyll~is a conservative code. Though the spatial resolution of \gkeyll~is less than \gdb's, the convergence of the radial ($k_x$) spectrum upon grid refinement~\cite{Bernard2019} and the similarities with \gdb~in quantities compared below suggests that the resolution is sufficient. It is still possible that metrics other than the $k_x$ spectrum would have shown greater variance. This was not explored exhaustively, and higher resolution simulations with the new, faster version of \gkeyll~may shed light on this point. Readers may recall, however, that it is longer wavelength modes that tend to drive most of the transport, and achieving fine-scale agreement between \gkeyll~and \gdb~may only affect the more intricate details of the turbulence.

\begin{figure}
\includegraphics[width=0.48\textwidth]{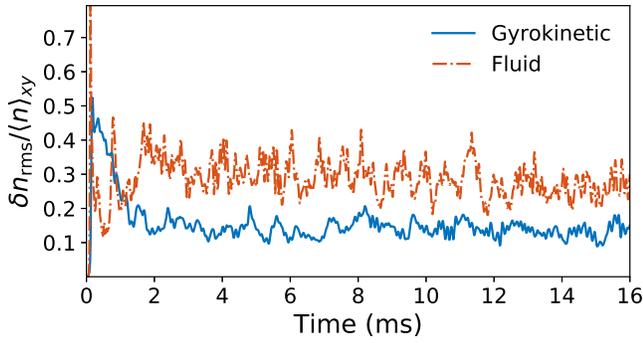}
\caption{\label{fig:nRMSrFluc} Relative root-mean-square of the density fluctuations in the $z=0$ plane as a function of time. The saturation current $\txtS{I}{sat}=n\sqrt{T_e}$ is used as a density proxy as is typically done with experimental probe data. This fluid simulation used $q_{\parallel s}\neq0$ and $2.63\SOfl$.}
\end{figure}

The strength of the turbulent fluctuations is however different in \gdb~and \gkeyll~at the mid-plane ($z=0$), at least as far as the relative root-mean-square (RMS) fluctuations in the saturation current ($\txtS{I}{sat}$) is concerned. Following experimental convention, instead of the density we use $\txtS{I}{sat}=n_e\sqrt{T_e}$ in computing density fluctuations, but for simplicity refer to it as $\delta n$ unless stated otherwise. Figure~\ref{fig:nRMSrFluc} provides a calculation of the $z=0$ $\delta n$ RMS fluctuation amplitude calculated using the instantaneous $y$-average, $\delta n = n-\avg{n}_y$, and normalized to $\avg{n}_{xy}$. We see that at $z=0$ and computing $\delta n$ using the instantaneous $y-$average yields a $\delta n_{\mathrm{rms}}/\left\langle n\right\rangle_{xy}$ that is twice as high in the fluid than in the gyrokinetic simulation. We will later see that this may be dominated by fluctuations levels in the high-field side of the fluid simulation, on which parallel heat-flux BCs have a significant impact. These are also considerably higher fluctuation levels than in previous simulations, using simpler fluid models, of Helimak~\cite{Li2009}.

\begin{figure}[h]
\includegraphics[width=0.49\textwidth]{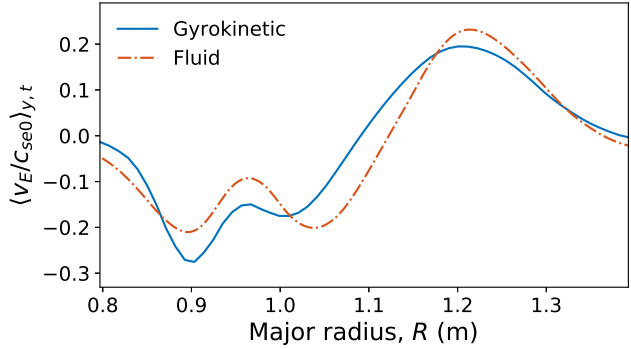}
\caption{\label{fig:vExB}Binormal component of the \ExB~drift normalized to the reference sound speed, averaged over $y$ and time.}
\end{figure}

Such qualitative differences arise throughout movies of the plasma density, yet these videos also reveal similarities in the binormal flows of both simulations. We obtained the time and $y$-averaged binormal component of the \ExB~drift velocity from both simulations and plot them in figure~\ref{fig:vExB}. These sub-sonic flows are particularly comparable on the low-field side where most of turbulence is located. The maximum $v_E$ is only 6\% higher in the fluid simulation and is located 3 cm farther out than the $R=1.2$ m location of the gyrokinetic peak $v_E$. An estimate of the experimental \ExB~profile using $\txtS{\phi}{exp} = \Lambda\txtS{T}{$e$,exp}/e$ leads to the conclusion that both the fluid and gyrokinetic simulation produce a $v_{E}$ that is quite different from that in the experiment~\cite{Bernard2019}. As explained in such previous work, there is an important vertical component to the \ExB~flow that can be larger than the vertical projection of the parallel sonic flows. Incorporating these effects would require a more accurate description of the geometry and is beyond the scope of this work.

\begin{figure}
\includegraphics[width=0.49\textwidth]{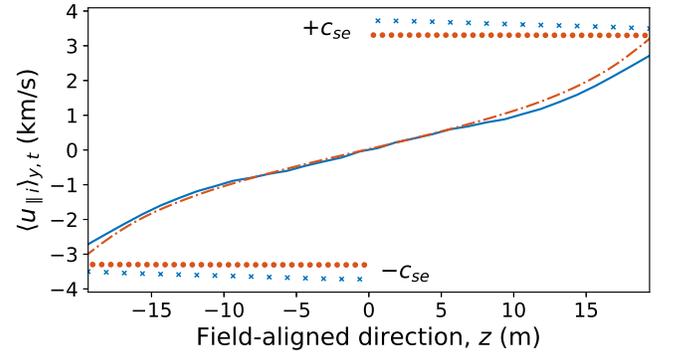}
\caption{\label{fig:vpiZ}$y$ and time averaged ion parallel velocity at $R=1.24$ m in the gyrokinetic (solid blue) and fluid (orange dash-dot) simulations. We also show the time and $y$ averaged zeroth-order sound speed (at the same $R$) in the gyrokinetic (blue crosses) and fluid (orange circles) calculations.}
\end{figure}

\begin{figure}
\includegraphics[width=0.49\textwidth]{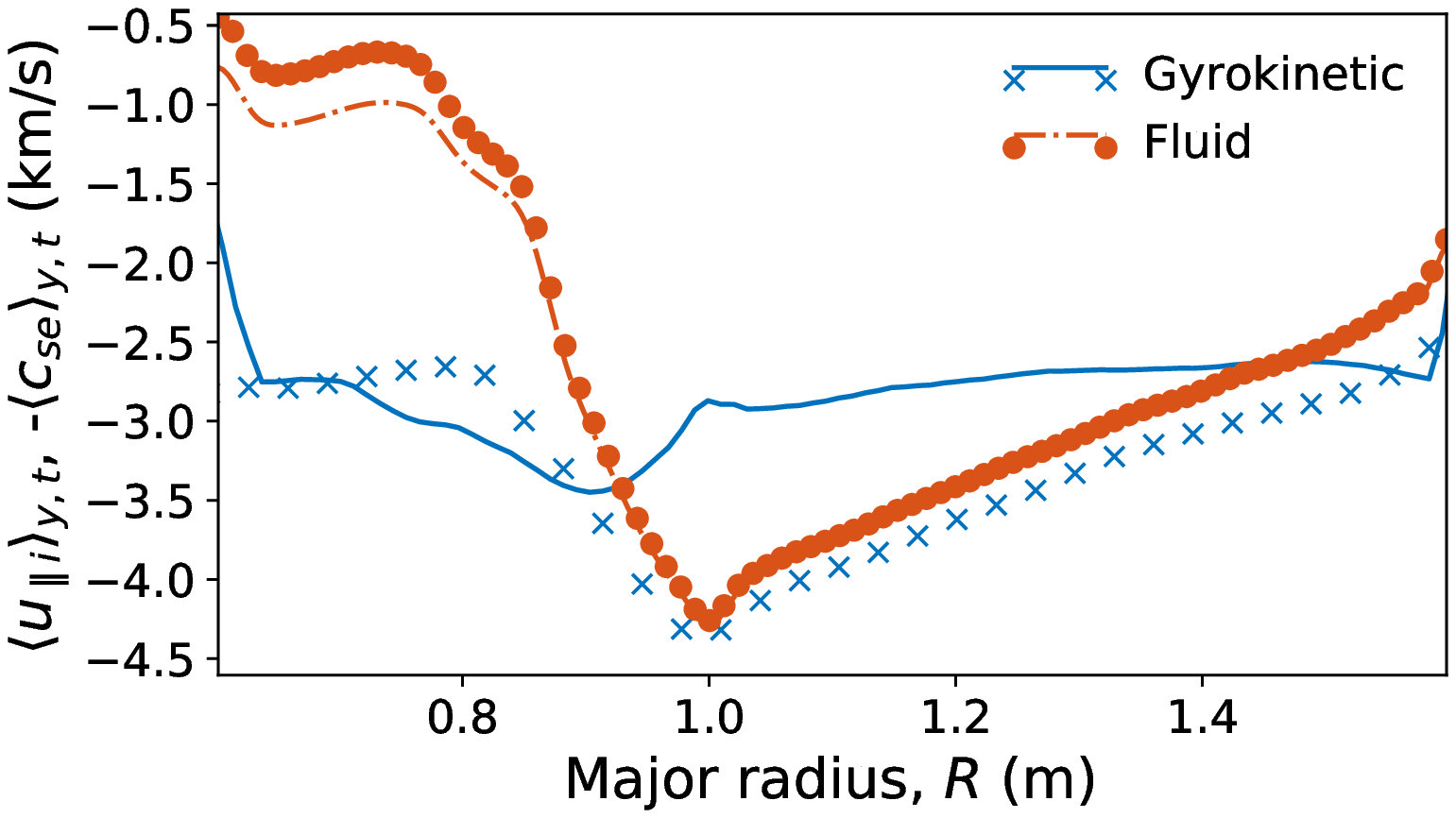}
\caption{\label{fig:vpiVSx}Ion parallel velocity, $y$ and time averaged, at the bottom sheath as a function of radius in the gyrokinetic (solid blue) and fluid (orange dash-dot) simulations. Also shown is the time and $y$ averaged zeroth-order sound speed in the gyrokinetic (blue crosses) and fluid (orange circles) calculations.}
\end{figure}

Continuing with the examination of flows, we compared parallel ion velocities from the \gdb~and \gkeyll~simulations by averaging $\vpi$ in $y$ and time at $R=1.24$ m (figure~\ref{fig:vpiZ}). Near the center of the $z$-domain, $\vpi$ is nearly equivalent in both models, but there are slight differences in gradients and more significant differences near the sheath. The larger $\upari$ gradient in the fluid simulation may seem suggestive a stronger particle outflow and thus a need for greater source rates, but as we now know the integrated particle flux out is similar for both codes (figure~\ref{fig:GammaNtt}). Ultimately the parallel losses are set by the exit value of the flux, which depends on a non-trivial density profile and the exit parallel velocity. The latter is forced to $\vpi=\pm c_s$ in \gdb, while \gkeyll's conducting sheath model does not enforce the Bohm criterion, which could explain the differences in $\vpi$ near the sheath. In this case, the gyrokinetic parallel ion flow at the sheath entrance was 23\% lower than the local $y$ and time averaged value of the sound speed ($T_i\ll T_e$ there).

A more complete picture is developed by considering the variation of $\upari$ across the radius of the machine. Figure~\ref{fig:vpiVSx} illustrates that departures from the Bohm criterion in the gyrokinetic simulation are even greater near the source region (compare solid blue line with blue crosses). The difference between the fluid $\upari$ and $\cse$ at $R<0.9$ m is a consequence of enforcing $\vpi=\pm c_s=\sqrt{(T_e+T_i)/m}$ and that $T_i>T_e$ in the high-field region. On the other side, in the low-field region where most of the plasma is found, \gdb's $\upari$ is almost consistently greater than that in \gkeyll. Therefore, the parallel transit $\Lc/(2\vpi)$ is slower in the gyrokinetic simulation and the plasma has more time to transport radially across field lines, in agreement with the average density profiles and turbulent snapshots presented above. 

\begin{figure}
\includegraphics[width=0.49\textwidth]{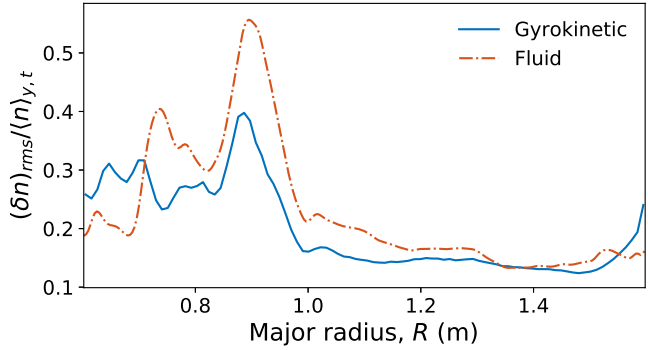}
\caption{\label{fig:nRMS}Relative root-mean-square fluctuation density profile for gyrokinetic (solid blue) and fluid (orange dash-dot) simulations.}
\end{figure}

Weaker parallel flows are generally associated with more cross-field transport. However, the gyrokinetic simulation shows lower turbulence levels than the fluid simulation as measured by the relative RMS fluctuations using the instantaneous $y$-average at $z=0$ (figure~\ref{fig:nRMSrFluc}). The presence of fine-scale structures in the $y$ direction of the fluid simulation seen in figure \ref{fig:snapshots} does not explain this discrepancy since lower $k_y$ modes tend to drive most of the transport. We, therefore, compared the $\delta n_{rms}$ radial profile in figure \ref{fig:nRMS} using $\delta n=n-\avg{n}_{y,t}$, i.e. the time and $y$ average instead of the instantaneous $y$-average alone. The peak relative fluctuation level occurs at the same location for both the fluid and gyrokinetic simulations, though it is 25\% higher in \gdb~and it occurs on the high-field side where many other discrepancies between the codes are seen (e.g., time-averaged profiles in figure~\ref{fig:profilesVsrc}). On the low-field side \gkeyll's relative $\delta n_{rms}$ is in fact slightly lower than \gdb's, so the more effective cross-field spreading of the gyrokinetic density is likely a more direct consequence of the difference in the flows.

\begin{figure}
\includegraphics[width=0.49\textwidth]{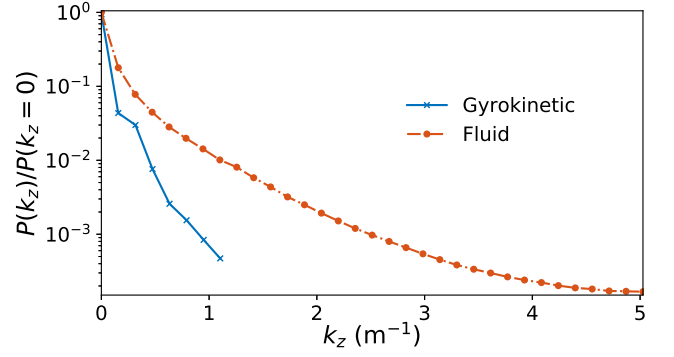}
\caption{\label{fig:parallelSpectra}Spectra of the density fluctuations along $z$, averaged over $y$ and time. The gyrokinetic spectra was also averaged over $x$, while the fluid data only averaged the spectra at the locations $R=0.83$ m, $R=1.0$ m, and $R=1.24$ m.}
\end{figure}

\begin{figure*}
\includegraphics[width=\textwidth]{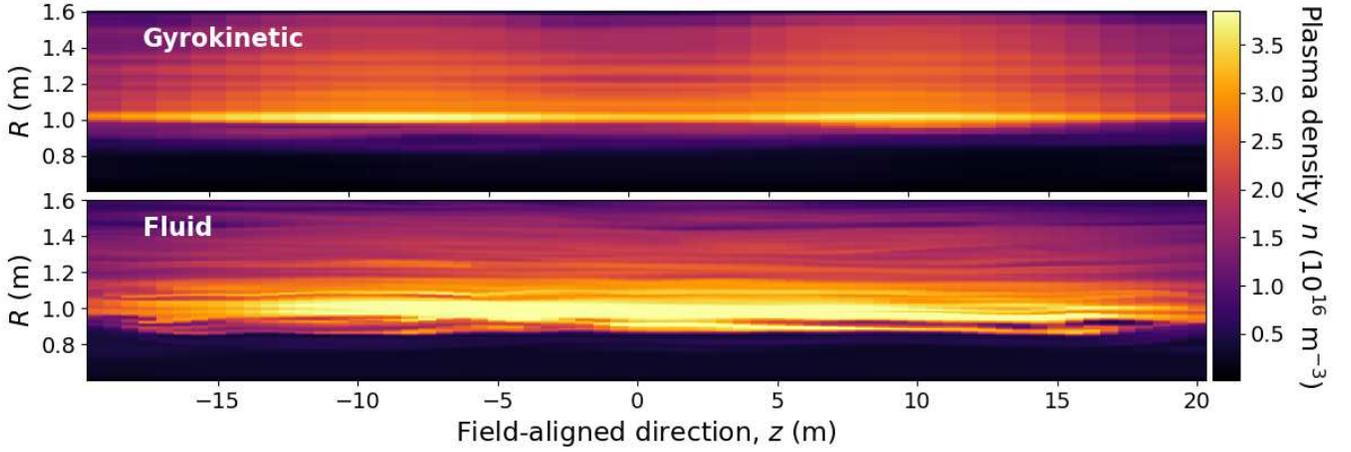}
\caption{\label{fig:xzSnapshot}Gyrokinetic (top) and fluid (bottom) plasma density snapshots at $y=0$ and $t=10$ ms. Colors set by extrema in gyrokinetic data.}
\end{figure*}

These RMS fluctuation profiles are nevertheless qualitatively similar: peaking in the high-field region and relatively flat in the outboard side. Fluctuations also have comparable qualities in the parallel direction. Take the power in the $k_z$ spectra of the density fluctuations in both simulations, for example (figure~\ref{fig:parallelSpectra}, using density fluctuations computed with $\delta n=n-\avg{n}_y$). The \gkeyll~spectrum was computed by Fourier transforming cell-average values. The $y$- and time-averaged spectra in figure~\ref{fig:parallelSpectra} both decay rapidly beyond $k_z=0$.  This $k_z\approx 0$ feature is characteristic of the interchange turbulent regime, which is predicted for the high field-line pitch angle used in these simulations~\cite{Ricci2010}. The lack of power in high $k_z$ modes is also observed in snapshots of the plasma density in $x-z$ at $t=10$ ms (figure~\ref{fig:xzSnapshot}), with both simulations showing little variation along field lines.

The fluid $k_z$ spectrum was higher than \gkeyll's for all finite $k_z$ modes, though we note that \gdb's spectra can be altered artificially by the use of additional numerical parallel diffusion. The \gdb~code is able to run with $\chi_\parallel=0$ but a centered finite-difference scheme without any upwinding can generate more $k_z\neq 0$ structures, significantly altering the parallel spectrum. Hence, small parallel diffusion terms were included, as indicated in appendix~\ref{sec:gdbNorm}. Finite parallel diffusion has stability benefits and is also used to regularize the $k_z$ spectrum, though it was not adjusted deliberately to match \gkeyll~results.


\begin{figure}
\includegraphics[width=0.49\textwidth]{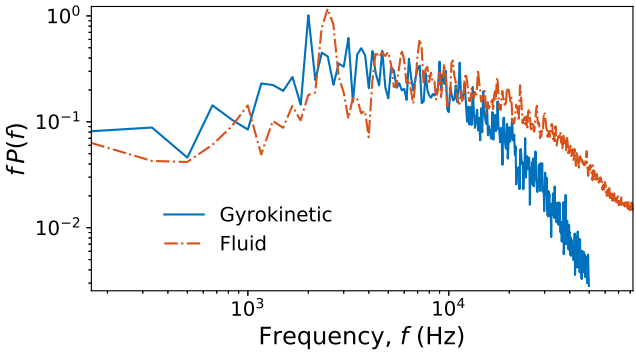}
\caption{\label{fig:fspectrum}Frequency spectrum of the density fluctuations of the gyrokinetic (solid blue) and fluid (orange dash-dot) simulations computed near the location of maximum $v_E$.}
\end{figure}

We also observed similarities in frequency-domain power distribution. As done in previous analysis of \gkeyll~data~\cite{Bernard2019}, we multiplied the time signal of the fluctuations $\delta n=n-\avg{n}_t$ by a Hann window
\begin{equation}
    \delta n_w(t_i) = \left[1-\cos\left(\frac{2\pi}{N_t-1}i\right)\right]\delta n(t_i)
\end{equation}
to account for the fact that the fluctuation data is not periodic at the first and last time frames, $t_0$ and $t_{N_t-1}$, respectively. Upon Fourier transforming this quantity to the frequency domain, we compute the normalized power spectra via
\begin{equation}
    P(f) = \frac{\avg{|\delta n_w(f)|^2}}{\sum_f (\Delta f)\avg{|\delta n_w(f)|^2}}.
\end{equation}
By using a 6 ms time signal, we resolved the frequency domain with a frequency spacing of $\Delta f=168$ Hz in \gkeyll~and $\Delta f=166$ Hz in \gdb. The frequency spectra (multiplied by the frequency) given in figure~\ref{fig:fspectrum} were computed at the location of maximum \ExB~drift in \gkeyll~($R=1.2$ m) and near the location of maximum \ExB~drift in \gdb~($R=1.24$ m). The fluid and gyrokinetic frequency spectra are comparable in the region where values are greatest, $1 - 10$ kHz. The spectra peak at slightly different frequencies, both of which are higher than the experimental peak~\cite{Bernard2019}. The spectrum magnitude for \gdb\ is larger than \gkeyll\ in the high frequency region. For example, at the highest frequency resolved by \gkeyll~(50 kHz), the power was an order of magnitude lower compared to the \gdb~spectrum. This may be associated with the rapidly-changing small-scale structures observed in \gdb, which could alter the spatial and temporal spectra, while only having a minimal effect on turbulent transport.

\begin{figure}
\includegraphics[width=0.49\textwidth]{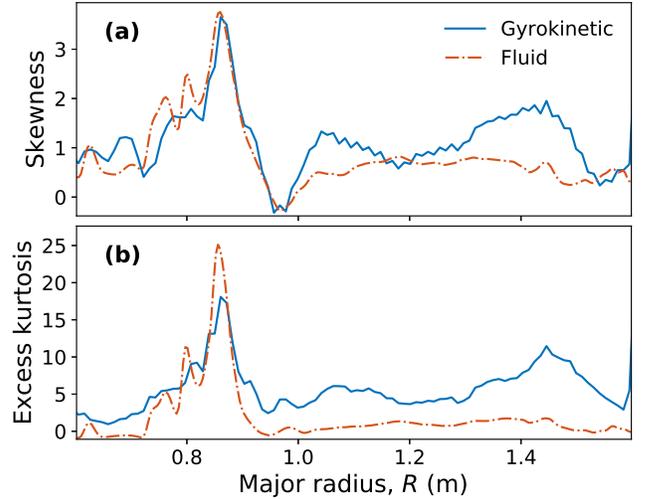}
\caption{\label{fig:skewKurt}(a) Skewness and (b) excess kurtosis of the fluctuations as a function of major radius in the gyrokinetic (solid blue) and fluid (orange dash-dot) simulations.}
\end{figure}

Finally, we probed statistical properties of the turbulence via the moments of the fluctuations' probability density function. The skewness and the excess kurtosis of the fluctuations as a function of radius are given in figure~\ref{fig:skewKurt}. Concurrent with the agreement in the location of the peak $\delta n_{rms}/\avg{n}_{y,t}$ near $R=0.9$ m (figure~\ref{fig:nRMS}), there is also agreement between the fluid and gyrokinetic simulations in the location of maximum skewness and excess kurtosis. There is a second peak in the fluid data at $R\approx 0.76$ m that is absent in the gyrokinetic simulation. Additionally, both skewness and kurtosis were consistently larger on the low-field side in \gkeyll, which corresponds to a flatter density profile in this region and is consistent with previous analyses of intermittent turbulence\cite{d2011convective}. 

\section{Additional discussion} \label{sec:discuss}

In this section we offer further commentary on aspects that are relevant to the physical and numerical facets of comparing \gdb~and \gkeyll, as well as to potential future comparisons between drift-reduced Braginskii and gyrokinetic models.

The comparison in this manuscript motivated the use of finite parallel heat-flux BCs in the fluid code. Although such BCs were necessary to achieve peak temperatures similar to that in the gyrokinetic simulation, they produced \gdb~profiles with a high-field side $T_e$ that was much smaller than that in \gkeyll. The smallest $T_e$ was produced by those finite heat-flux BCs not accounting for convective and frictional contributions ($q_{\parallel s}\neq0$). In the future it may be possible to achieve greater agreement between the high-field side $T_e$ produced by the two codes by including the spatial variation of $\kappa_{\parallel}^s$ (supported in \gdb~but not used here), which would reduce the amount of heat extracted in regions where $T_e<T_{e0}$.

Closing the gap between \gdb~and \gkeyll~could necessitate enhancements to the latter as well. A reader may expect that a Braginskii model, as moments of the kinetic equation, would produce similar results to those obtained by solving the (long wavelength) gyrokinetic equation for this highly collisional plasma. However, aside from the drift-reduction assumed in \gdb, the Braginskii equations used here cannot be derived by taking moments of \gkeyll's gyrokinetic equation~\ref{eq:fDot}. One conflict is that finite Larmor radius (FLR) effects are only partially incorporated in the \gkeyll~model via the Poisson equation (\ref{eq:poisson}) but not included in the gyrokinetic equation (\ref{eq:fDot}). It would be necessary to include FLR effects in order to derive the correct gyroviscous terms from moments of the gyrokinetic equations. Another obstacle is \gkeyll's model collision operator having transport coefficients different to those in the Braginskii model. The viscosity coefficient arising from the Dougherty operator, for example, is $\eta_{0,\text{Dougherty}}^s=0.5 n T_s/(\nu_{ss}+\nu_{sr})$~\cite{Anderson2007}. On the other hand the Braginskii transport coefficients are computed from the exact linearized Landau collision integral, and give the viscosities $\eta_0^i=0.96nT_i/\nu_{ii}$ and $\eta_0^e=0.73nT_e/\nu_{ei}$. One could change the Braginskii viscosity, and other transport coefficients, to better approximate \gkeyll's. In fact, previous fluid simulations of the Helimak either used an artificial value of $\eta_0^e$ or neglected it altogether. Including an artificial value of $\eta_0^e$ is more important when using reduced mass ratios, and it can significantly alter the density and current profiles~\cite{Francisquez2018a}. However, it is still unclear to what extent matching transport coefficients is crucial to the accurate simulation of this system. A safer course of action is to implement a more accurate collision operator, several of which are being developed that more closely approach Braginskii transport coefficients in the highly collisional limit~\cite{Pan2019,Sugama2019,Jorge2019}.

The collision operator, through its higher moments, also provides the dissipative channel in the kinetic system. This leads to, for example, collisional drift terms that have been proposed for inclusion in fluid models~\cite{Madsen2016}. These terms are typically not considered as they are thought to be small compared to the artificial diffusion ($\propto\mathcal{D}$) required by the numerical methods used in Braginskii codes. Those diffusion terms can still impact the properties of the turbulence, even though at the levels reported here they did not account for a big portion of \gdb's particle balance. In the future exactly conservative formulations of fluid equations that obviate the need for these hyperdiffusion terms may serve as a more reliable approach~\cite{Halpern2018}.

In this sheath-dominated regime, the collisional refinements are likely secondary to the influence of the sheath BCs. As explained in section~\ref{sec:models}, the parallel BCs in both codes are not equivalent. \gkeyll's innovative conducting sheath boundary conditions~\cite{Shi2017}, have been successfully used to model LAPD, Helimak, and NSTX. Despite their adoption by kinetic codes, the \gkeyll~Helimak simulations demonstrate that these BCs do not satisfy the Bohm sheath criterion. On the other hand, fluid codes almost universally impose the Bohm criterion, either $\vpi=\pm c_s$ or $|\vpi|=\geq c_s$. It would be useful to know how the kinetic conducting BCs can be modified in order to satisfy the Bohm sheath criterion, perhaps by developing an improved rule for the reflection of the electrons. However, it is unknown whether Bohm sheath BCs are the correct choice for all simulations of laboratory plasmas. Derivation of this condition, for example, assumes ambipolar flows~\cite{Stangeby2000}, but a significant fraction of non-ambipolarity has been measured in the tokamak scrape-off-layers~\cite{Dejarnac2015}. Therefore, more experimental diagnostics at the sheath will likely prove helpful in exploring improved parallel BCs for gyrokinetic and fluid models.

Beyond collision operators and parallel BCs, there are other interesting enhancements that can be pursued with both codes.
We previously mentioned that there are better descriptions of the geometry that could be implemented, and new versions of \gkeyll~and \gdb~already contain these capabilities. The new version of \gkeyll, and the flux-coordinate independent approach implemented in \gdb~tokamak simulations, can be used to incorporate shear and as well as the vertical component of the \ExB~velocity. Both codes can also run without the Boussinesq approximation. In tokamak simulations this sophistication did not always alter the results significantly~\cite{Francisquez2017,Ross2018}, but no exhaustive scans of parameter space have been performed. In the few Helimak simulations we have performed, we note that incorporating the spatial variation $n(x)/B^2(x)$ in the ion polarization can add a modest change to the perpendicular profiles and very drastic changes to the parallel current profiles~\cite{Francisquez2018a}.
The aforementioned enhancements may however turn out to be minor when confronted with the high levels of input power radiated away in the Helimak (>90\%~\cite{Bernard2019}); including ionization, radiation cooling of electrons and charge exchange will be essential for fully predictive simulations.

\begin{figure}
\includegraphics[width=0.49\textwidth]{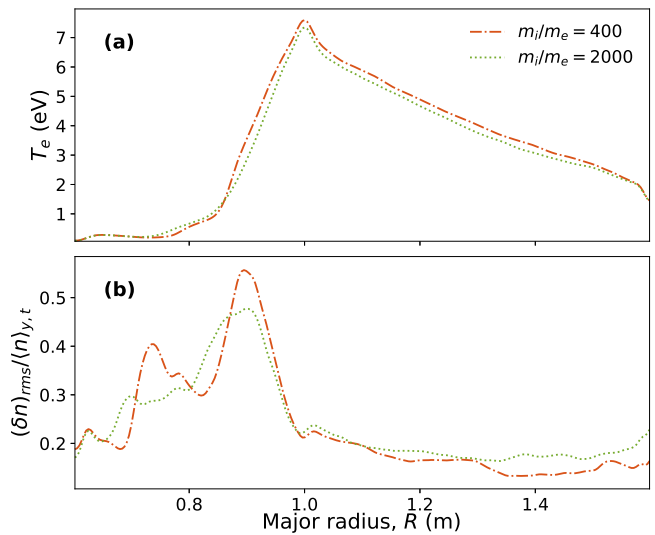}
\caption{\label{fig:TeVpiMrat}(a) Radial electron temperature and (b) root-mean-squared (RMS) fluctuation profiles at the bottom sheath for \gdb~simulations with $m_i/m_e=400$ (orange dash-dot) and $m_i/m_e=2000$ (dotted green). These were produced by the same procedure as figures~\ref{fig:profiles} and~\ref{fig:nRMS}.}
\end{figure}

One improvement that we pursued and present here is the use of a higher mass ratio. Since the \gdb~simulation is more than 10 times cheaper we were able to run it with $m_i/m_e=2000$ with a negligible increase in cost. This is still much less than the true Argon $m_i/m_e=7.33\times10^4$. These lighter electrons resulted in a small reduction of the electron temperature profile (figure \ref{fig:TeVpiMrat}a) and a relatively small correction to the RMS fluctuation profile (figure~\ref{fig:TeVpiMrat}b). Additional results from these simulations can be found in the supplemental materials.

\section{Summary and closing remarks} \label{sec:conclusion}

We have presented a detailed comparison of plasma turbulence simulations on open-field lines produced by the fluid code \gdb~and the (long wavelength) gyrokinetic code \gkeyll. An examination of the radial profiles prompted the implementation of two finite heat-flux BCs ($q_{\parallel s}\neq0$ and $q_{\parallel e}^{\mathrm{tot}}\neq0$) in \gdb; these cause important modifications of the temperature profiles. The differences in fluid and gyrokinetic global profiles also motivated the use of larger source amplitudes in \gdb~than those matching the volume integral of the sources \gkeyll. This was in part necessary due to the lack of particle conservation in \gdb, and an examination of this feature highlighted the need to account for missing Jacobian factors and for the inhomogeneity of the magnetic field in the \ExB~nonlinearity.

Comparing the gyrokinetic simulation to the $q_{\parallel s}\neq0$ fluid simulation with  sources increased 2.1875 times and without the $B=B_{(R)}$ accounted for in the \ExB~nonlinearity, we observed differences in the high-$f$ spectrum, skewness and kurtosis of density fluctuations, and the density gradients. It was also shown that \gkeyll's conducting sheath BCs do not satisfy the Bohm sheath criterion, while \gdb's BCs impose such condition. Sheath BCs can indirectly affect the turbulence and the cross-field spreading of the density profile by altering the particle transit time. At the same time, several other quantities exhibited relative agreement. For example, turbulent structures were qualitatively similar in $x$-$y$ snapshots, and \ExB~flow and turbulent fluctuation profiles were qualitatively, as well as in some cases quantitatively, close.

Future \gdb-\gkeyll~comparisons will hopefully account for Jacobian factors and inhomogeneities were needed, as well as employ improvements to conservation and a spatially varying collisionality. Ultimately however, it will be crucial to accompany these fluid-kinetic comparisons with experimental validations. If one compares the results presented here with experimental data~\cite{Bernard2019} it becomes clear that both codes require improvements. The gyrokinetic model produced density fluctuations with a skewness profile and a frequency spectrum that is closer to those observed experimentally than those obtained with the fluid code. But the average density, temperature, and $\delta n_{rms}$ radial profiles in both codes have notable departures from experimental data. In the future we will use and implement improvements to various features in both codes. This task is partially aided by the reduced cost of \gdb~simulations; adding and testing new features in the fluid model can help determine which improvements should be prioritized in the gyrokinetic model in order to use resources more efficiently. The new, faster version of \gkeyll~also allows for more rapid development and testing.

Finally, a key inquiry of these fluid-kinetic comparisons is whether the fluid or the gyrokinetic model is more appropriate for modeling this and other SOL-like, open-field line systems. The model in \gkeyll~may at present be incomplete, but the gyrokinetic system it is developing towards is free of certain limitations inherent to the a moment, collisional closure as Braginskii's. As such it will eventually provide superior accuracy. But both \gkeyll~and \gdb~are not fully developed; given the maturity of these models and the data presented here we cannot provide a universal recommendation on which kind of code is more suitable. In general we can remind the reader that a gyrokinetic code may be better suited to model a lower-collisionality regime and to study kinetic effects. A fluid code is more computationally affordable for modeling a high-collisionality regime and is still able to capture many features of the gyrokinetic simulations. It is clear that improvements are needed in both models, though the general agreement between the two is encouraging for both sides. This work represents a starting point for future comparisons of fluid and kinetic models, including their respective strengths and weaknesses, which will be essential in the effort to achieve predictive modeling of advanced fusion devices.

\begin{acknowledgments}
We thank Paolo Ricci for his insight into the early simulations of the Texas Helimak and Eric Shi for his role in the development of \gkeyll. Discussion with Ken Gentle and Edward Taylor was also key to understanding the experimental configuration and establishing the numerical setup. We thank Darin Ernst and the rest of the \gkeyll~team for their input and helpful suggestions. The simulations presented here carried out at the Texas Advanced Computing Center, the Dartmouth Discovery cluster, and MIT-PSFC's partition of the Engaging cluster at the MGHPCC facility (\url{www.mghpcc.org}, funded by DoE grant number DE-FG02-91-ER54109), so we wish to thank the support teams at these facilities for their attention and great work. M.F., B.Z. and B.N.R were supported by the U.S. Department of Energy (DOE) grant DOE-SC-0010508. M.F. is currently supported by DOE contract DE-FC02-08ER54966. T.N.B. was supported by DOE contract DE-FG02-04ER-54742, through the Institute of Fusion Studies at the University of Texas at Austin, and is currently supported by DOE contract DE-FG02-95ER54309. B.Z. is now supported by the Tokamak Disruption Study SciDAC Project and by DOE contract DE-AC52-07NA27344 through the Lawrence Livermore National Laboratory. A.H. and G.W.H. are supported by the High-Fidelity Boundary Plasma Simulation SciDAC Project, part of the DOE Scientific Discovery Through Advanced Computing (SciDAC) program, through DOE contract DE-AC02-09CH11466 for the Princeton Plasma Physics Laboratory. A.H. is also supported by the Air Force Office of Scientific Research under contract FA9550-15-1-0193. The reduced data used to produce the figures in this work is available upon request, and input files for the simulations can also be furnished should the raw data need regeneration.
\end{acknowledgments}

\appendix

\section{\gdb~normalization} \label{sec:gdbNorm}

The two-fluid equations~\ref{eq:nSrc}-\ref{eq:TiDot} are solved in the following normalized form:
\begin{align} \label{eq:ldenDotNorm}
\d{^e\ln n}{t} &= -\div{\v{v}_{\perp}}-\epv \delpar{\vpe}+\frac{S_n}{n}+\mathcal{D}_{\ln n}
\end{align}
\begin{align} \label{eq:wDotNorm}
\pd{\gvort}{t} &= \ncurv{p_e}+\tau\ncurv{p_i}+\frac{\epv}{\alphad\epa}\delpar{\cur}-\ncurv{G_i} \nonumber \\
&\quad-\div{\left\lbrace\frac{n}{B^3}\left[\phi,\v{h}\right]+\sqrt{\tau}\epv\frac{n}{B^2}\vpi\delpar{\v{h}}\right\rbrace} + \mathcal{D}_{\gvort}
\end{align}
\begin{align} \label{eq:vpeDotNorm}
\d{^e\vpe}{t} &= \frac{m_i}{m_e}\epv\left(\frac{\delpar{\phi}}{\alphad}-\frac{\delpar{p_e}}{n}-0.71\delpar{T_e}+4\frac{\delpar{G_e}}{n}\right) \nonumber \\
&\quad+\eta\cur+\epa\alphad T_e\ncurv{\vpe}-\frac{\vpe}{n}S_n+\mathcal{D}_{\vpe},
\end{align}
\begin{align} \label{eq:vpiDotNorm}
\d{^i\vpi}{t} &= -\frac{\epv}{\sqrt{\tau}}\left(\frac{\delpar{\phi}}{\alphad}+\tau\frac{\delpar{p_i}}{n}-0.71\delpar{T_e}-4\frac{\delpar{G_i}}{n}\right) \nonumber \\
&\quad-\frac{m_e}{m_i}\frac{\eta}{\sqrt{\tau}}\cur-\epa\tau\alphad T_i\ncurv{\vpi}-\frac{\vpi}{n}S_n+\mathcal{D}_{\vpi}
\end{align}
\begin{align} \label{eq:lTeDotNorm}
\d{^e\ln T_e}{t} &= \frac{5}{3}\epa\alphad\ncurv{T_e}+\frac{1}{p_e}\delpar{\kappa^e\delpar{T_e}}+\frac{2}{3}\left[ -\div{\v{v}_{\perp}} \right. \nonumber \\
&\left.\quad-\epv \delpar{\vpe}+\frac{0.71\epv}{n}\left(\delpar{\cur}-\cur\delpar{\ln T_e}\right) \right. \nonumber \\
&\left.\quad+\frac{S_{E,e}}{p_e}-\frac{3}{2}\frac{S_n}{n}\right]+\mathcal{D}_{\ln T_e}
\end{align}
\begin{align} \label{eq:lTiDotNorm}
\d{^i\ln T_i}{t} &= -\frac{5}{3}\tau\epa\alphad\ncurv{T_i}+\frac{1}{p_i}\delpar{\kappa^i\delpar{T_i}}+\frac{2}{3}\left(-\div{\v{v}_{\perp}} \right. \nonumber \\
&\left.\quad-\sqrt{\tau}\epv\delpar{\vpi}+\epv\frac{\delpar{\cur}}{n}+\frac{S_{E,i}}{p_i}-\frac{3}{2}\frac{S_n}{n}\right)+\mathcal{D}_{\ln T_i}
\end{align}
where we used $\div{\v{v}_{\perp}}=\epa\left[\ncurv{\phi}-\alphad \ncurv{p_e}/n\right]$, $\v{h}=\grad\phi+\tau\alphad\left(\grad{p_i}\right)/n$, $\gvort=\div{n\v{h}/B^2}$ is the generalized vorticity, and $\cur=n\left(\sqrt{\tau}\vpi-\vpe\right)$ is the normalized current. The dynamic variables in physical units can be retrieved as follows: $\txtS{n}{phys}=\nRef n$, $\txtSb{T}{\alpha}{,phys}=\txtSb{T}{\alpha}{,ref}T_{\alpha}$, $\txtS{\phi}{phys}=\BRef a^2\phi/(c\tRef))$, $\txtSb{v}{\parallel \alpha}{,phys}=\txtSb{v}{\parallel\alpha}{,ref}v_{\parallel\alpha}$, $\txtSb{j}{\parallel}{phys}=e\nRef\vpeRef\cur$. The dimensionless magnetic field and major radius are defined by $\txtS{B}{phys}=\BRef B$ and $\txtS{R}{phys}=\RRef R$.
The reference values were $\nRef=n_{e0}$, $\BRef=\BO$, $\RRef=R_0$, $\txtSb{T}{s}{,ref}=T_{s0}$, $\txtSb{u}{\parallel s}{,ref}=\txtSb{c}{ss}{,ref}=\sqrt{\txtSb{T}{s}{,ref}/m_s}$.
Perpendicular lengths are normalized to the machine's width $a=1$ m, and parallel ones to the plasma center major radius $R_0$. The reference time is the interchange-like timescale $\tRef=\sqrt{aR_0/2}/\cseRef$, which for the reference parameters in this work gives $\tRef=1.5145\times10^{-4}$ s.
The normalized transport coefficients are $\epgi=0.08\tau\tauiRef/\tRef$, $\epge=0.73\taueRef/(12\tRef)$, $\eta=0.51\tRef/\taueRef$, $\kappae=3.2[2\tRef/(3\nRef\LparRef^2)]\taueRef\nRef\TeRef/m_e$ and $\kappai=3.9[2\tRef/(3\nRef\LparRef^2)]\tauiRef\nRef\TiRef/m_i$.
Here $\txtSb{\tau}{s}{,ref}$ refers to the collisional period~\cite{Huba2013}. We also employ the dimensionless variables $\epa=2a/R_0$, $\epv=\cseRef\tRef/R_0$, $\alphad=\cseRef^2\tRef/(\txtSb{\Omega}{i}{,ref} a^2)$. The normalized functions arising from the gyroviscous stress tensor are now
\begin{align}
G_i &= \epgi\left[4\sqrt{\tau}\epv\delpar{\vpi}+\epa\left(\ncurv{\phi}+\tau\alphad\frac{\ncurv{p_i}}{n}\right)\right], \\
G_e &= \epge\left[4\epv\delpar{\vpe}+\epa\left(\ncurv{\phi}-\alphad\frac{\ncurv{p_e}}{n}\right)\right].
\end{align}
The normalized form of the time rate of change is
\begin{align}
\d{^e F}{t} &= \pd{F}{t}+\frac{1}{B}\left[\phi,F\right]+\epv\vpe\delpar{F}, \\
\d{^i G}{t} &= \pd{G}{t}+\frac{1}{B}\left[\phi,G\right]+\sqrt{\tau}\epv\vpi\delpar{G},
\end{align}
and the normalized curvature operator is $\ncurv{F}=\partial F/\partial y$. We also require the appropriate normalized form of the finite heat flux BCs in equation~\ref{eq:heatFlux}. This is
\begin{eqnal}
\delpar{\ln T_e} &= \mp \frac{\sqrt{\epa}}{3\kappa^e}\,\gamma_e n \sqrt{T_e+\tau T_i}\,\exp\left[\Lambda-\max\left(\frac{\phi}{\alphad T_e},0\right)\right], \\
\delpar{\ln T_i} &= \mp \frac{\sqrt{\epa}}{3\kappa^i}\,\gamma_i n \sqrt{T_e+\tau T_i},
\end{eqnal}
where the upper (lower) sign is used at the top (bottom) sheath. The heat transmission coefficients are $\gamma_i=2.5\tau T_i/T_e$ and $\gamma_e=2+\left|\phi/(\alphad T_e)\right|$. On the other hand, the finite electron heat-flux BC that takes into account the convective and frictional component of the heat-flux (equation~\ref{eq:qeTot}) is implemented as
\begin{eqnal}
&\delpar{\ln T_e} = \mp\, n \sqrt{T_e+\tau T_i}\,\left\lbrace 0.71 \right.\\
&\left.\quad+ \frac{\sqrt{\epa}}{3\kappa^e}\,\left(\gamma_e-\frac{5}{2}-0.71\right)\,\exp\left[\Lambda-\max\left(\frac{\phi}{\alphad T_e},0\right)\right]\right\rbrace.
\end{eqnal}
Finally, unless stated otherwise the normalized diffusion coefficients were
\begin{center}
\begin{tabular}{c|c|c|c}
 & $\chi_x$ ($10^{-17}$) & $\chi_y$ ($10^{-18}$) & $\chi_\parallel$ \\
 \hline
$\mathcal{D}_{\ln n}$ & $0.30686$ & $0.34384$ & 0.30572 \\
$\mathcal{D}_{\ln T_e}$ & $0.30686$ & $0.34384$ & 0 \\
$\mathcal{D}_{\ln T_i}$ & $0.30686$ & $0.34384$ & 0 \\
$\mathcal{D}_{\vpi}$ & $0.30686$ & $0.34384$ & 0.030572 \\
$\mathcal{D}_{\gvort}$ & $6.5770$ & $7.3697$ & 0.30572 \\
$\mathcal{D}_{\vpe}$ & $1.4206$ & $1.5919$ & 0.30572.
\end{tabular}
\end{center}

\nocite{*}
\bibliography{main.bib}

\providecommand{\noopsort}[1]{}\providecommand{\singleletter}[1]{#1}%
\begin{thebibliography}{49}%
\makeatletter
\providecommand \@ifxundefined [1]{%
 \@ifx{#1\undefined}
}%
\providecommand \@ifnum [1]{%
 \ifnum #1\expandafter \@firstoftwo
 \else \expandafter \@secondoftwo
 \fi
}%
\providecommand \@ifx [1]{%
 \ifx #1\expandafter \@firstoftwo
 \else \expandafter \@secondoftwo
 \fi
}%
\providecommand \natexlab [1]{#1}%
\providecommand \enquote  [1]{``#1''}%
\providecommand \bibnamefont  [1]{#1}%
\providecommand \bibfnamefont [1]{#1}%
\providecommand \citenamefont [1]{#1}%
\providecommand \href@noop [0]{\@secondoftwo}%
\providecommand \href [0]{\begingroup \@sanitize@url \@href}%
\providecommand \@href[1]{\@@startlink{#1}\@@href}%
\providecommand \@@href[1]{\endgroup#1\@@endlink}%
\providecommand \@sanitize@url [0]{\catcode `\\12\catcode `\$12\catcode
  `\&12\catcode `\#12\catcode `\^12\catcode `\_12\catcode `\%12\relax}%
\providecommand \@@startlink[1]{}%
\providecommand \@@endlink[0]{}%
\providecommand \url  [0]{\begingroup\@sanitize@url \@url }%
\providecommand \@url [1]{\endgroup\@href {#1}{\urlprefix }}%
\providecommand \urlprefix  [0]{URL }%
\providecommand \Eprint [0]{\href }%
\providecommand \doibase [0]{http://dx.doi.org/}%
\providecommand \selectlanguage [0]{\@gobble}%
\providecommand \bibinfo  [0]{\@secondoftwo}%
\providecommand \bibfield  [0]{\@secondoftwo}%
\providecommand \translation [1]{[#1]}%
\providecommand \BibitemOpen [0]{}%
\providecommand \bibitemStop [0]{}%
\providecommand \bibitemNoStop [0]{.\EOS\space}%
\providecommand \EOS [0]{\spacefactor3000\relax}%
\providecommand \BibitemShut  [1]{\csname bibitem#1\endcsname}%
\let\auto@bib@innerbib\@empty
\bibitem [{\citenamefont {Kotschenreuther}\ \emph {et~al.}(1996)\citenamefont
  {Kotschenreuther}, \citenamefont {Dorland}, \citenamefont {Liu},
  \citenamefont {Hammett}, \citenamefont {Beer}, \citenamefont {Smith},
  \citenamefont {Bondeson},\ and\ \citenamefont
  {Cowley}}]{Kotschenreuther1996}%
  \BibitemOpen
  \bibfield  {author} {\bibinfo {author} {\bibfnamefont {M.}~\bibnamefont
  {Kotschenreuther}}, \bibinfo {author} {\bibfnamefont {W.}~\bibnamefont
  {Dorland}}, \bibinfo {author} {\bibfnamefont {Q.~P.}\ \bibnamefont {Liu}},
  \bibinfo {author} {\bibfnamefont {G.~W.}\ \bibnamefont {Hammett}}, \bibinfo
  {author} {\bibfnamefont {M.~A.}\ \bibnamefont {Beer}}, \bibinfo {author}
  {\bibfnamefont {S.~A.}\ \bibnamefont {Smith}}, \bibinfo {author}
  {\bibfnamefont {A.}~\bibnamefont {Bondeson}}, \ and\ \bibinfo {author}
  {\bibfnamefont {S.~C.}\ \bibnamefont {Cowley}},\ }\bibfield  {title}
  {\enquote {\bibinfo {title} {{First principles calculations of tokamak energy
  transport}},}\ }in\ \href@noop {} {\emph {\bibinfo {booktitle} {Proceedings
  of the 16th International Conference on Fusion Energy}}}\ (\bibinfo {address}
  {Montreal},\ \bibinfo {year} {1996})\ pp.\ \bibinfo {pages}
  {371--383}\BibitemShut {NoStop}%
\bibitem [{\citenamefont {Stangeby}(2000)}]{Stangeby2000}%
  \BibitemOpen
  \bibfield  {author} {\bibinfo {author} {\bibfnamefont {P.}~\bibnamefont
  {Stangeby}},\ }\href@noop {} {\emph {\bibinfo {title} {The Plasma Boundary of
  Magnetic Fusion Devices}}},\ Series in Plasma Physics and Fluid Dynamics\
  (\bibinfo  {publisher} {Taylor \& Francis},\ \bibinfo {year}
  {2000})\BibitemShut {NoStop}%
\bibitem [{\citenamefont {Myra}\ \emph {et~al.}(6)\citenamefont {Myra},
  \citenamefont {D'Ippolito}, \citenamefont {Stotler}, \citenamefont {Zweben},
  \citenamefont {Leblanc}, \citenamefont {Menard}, \citenamefont {Maqueda},\
  and\ \citenamefont {Boedo}}]{Myra2006}%
  \BibitemOpen
  \bibfield  {author} {\bibinfo {author} {\bibfnamefont {J.~R.}\ \bibnamefont
  {Myra}}, \bibinfo {author} {\bibfnamefont {D.~A.}\ \bibnamefont
  {D'Ippolito}}, \bibinfo {author} {\bibfnamefont {D.~P.}\ \bibnamefont
  {Stotler}}, \bibinfo {author} {\bibfnamefont {S.~J.}\ \bibnamefont {Zweben}},
  \bibinfo {author} {\bibfnamefont {B.~P.}\ \bibnamefont {Leblanc}}, \bibinfo
  {author} {\bibfnamefont {J.~E.}\ \bibnamefont {Menard}}, \bibinfo {author}
  {\bibfnamefont {R.~J.}\ \bibnamefont {Maqueda}}, \ and\ \bibinfo {author}
  {\bibfnamefont {J.}~\bibnamefont {Boedo}},\ }\bibfield  {title} {\enquote
  {\bibinfo {title} {{Blob birth and transport in the tokamak edge plasma:
  Analysis of imaging data}},}\ }\href {\doibase 10.1063/1.2355668} {\bibfield
  {journal} {\bibinfo  {journal} {Phys. Plasmas2}\ }\textbf {\bibinfo {volume}
  {13}},\ \bibinfo {pages} {092509} (\bibinfo {year} {6})}\BibitemShut
  {NoStop}%
\bibitem [{\citenamefont {Pitts}\ \emph {et~al.}(2007)\citenamefont {Pitts},
  \citenamefont {Andrew}, \citenamefont {Arnoux}, \citenamefont {Eich},
  \citenamefont {Fundamenski}, \citenamefont {Huber}, \citenamefont {Silva},
  \citenamefont {Tskhakaya},\ and\ \citenamefont {Contributors}}]{Pitts2007}%
  \BibitemOpen
  \bibfield  {author} {\bibinfo {author} {\bibfnamefont {R.~A.}\ \bibnamefont
  {Pitts}}, \bibinfo {author} {\bibfnamefont {P.}~\bibnamefont {Andrew}},
  \bibinfo {author} {\bibfnamefont {G.}~\bibnamefont {Arnoux}}, \bibinfo
  {author} {\bibfnamefont {T.}~\bibnamefont {Eich}}, \bibinfo {author}
  {\bibfnamefont {W.}~\bibnamefont {Fundamenski}}, \bibinfo {author}
  {\bibfnamefont {A.}~\bibnamefont {Huber}}, \bibinfo {author} {\bibfnamefont
  {C.}~\bibnamefont {Silva}}, \bibinfo {author} {\bibfnamefont
  {D.}~\bibnamefont {Tskhakaya}}, \ and\ \bibinfo {author} {\bibfnamefont
  {J.~E.}\ \bibnamefont {Contributors}},\ }\bibfield  {title} {\enquote
  {\bibinfo {title} {{ELM transport in the JET scrape-off layer}},}\ }\href
  {\doibase 10.1088/0029-5515/47/11/005} {\bibfield  {journal} {\bibinfo
  {journal} {Nucl. Fusion}\ }\textbf {\bibinfo {volume} {47}},\ \bibinfo
  {pages} {1437--1448} (\bibinfo {year} {2007})}\BibitemShut {NoStop}%
\bibitem [{\citenamefont {{National Academies of Sciences, Engineering, and
  Medicine}}(2019)}]{NAP2019}%
  \BibitemOpen
  \bibfield  {author} {\bibinfo {author} {\bibnamefont {{National Academies of
  Sciences, Engineering, and Medicine}}},\ }\href {\doibase 10.17226/25331}
  {\emph {\bibinfo {title} {{Final Report of the Committee on a Strategic Plan
  for U.S. Burning Plasma Research}}}}\ (\bibinfo  {publisher} {{The National
  Academies Press}},\ \bibinfo {address} {{Washington, DC}},\ \bibinfo {year}
  {2019})\BibitemShut {NoStop}%
\bibitem [{\citenamefont {Ricci}, \citenamefont {Rogers},\ and\ \citenamefont
  {Brunner}(2008)}]{Ricci2008}%
  \BibitemOpen
  \bibfield  {author} {\bibinfo {author} {\bibfnamefont {P.}~\bibnamefont
  {Ricci}}, \bibinfo {author} {\bibfnamefont {B.~N.}\ \bibnamefont {Rogers}}, \
  and\ \bibinfo {author} {\bibfnamefont {S.}~\bibnamefont {Brunner}},\
  }\bibfield  {title} {\enquote {\bibinfo {title} {{High- and low-confinement
  modes in simple magnetized toroidal plasmas}},}\ }\href {\doibase
  10.1103/PhysRevLett.100.225002} {\bibfield  {journal} {\bibinfo  {journal}
  {Phys. Rev. Lett.}\ }\textbf {\bibinfo {volume} {100}},\ \bibinfo {pages}
  {6--9} (\bibinfo {year} {2008})}\BibitemShut {NoStop}%
\bibitem [{\citenamefont {Li}\ \emph {et~al.}(2009)\citenamefont {Li},
  \citenamefont {Rogers}, \citenamefont {Ricci},\ and\ \citenamefont
  {Gentle}}]{Li2009}%
  \BibitemOpen
  \bibfield  {author} {\bibinfo {author} {\bibfnamefont {B.}~\bibnamefont
  {Li}}, \bibinfo {author} {\bibfnamefont {B.~N.}\ \bibnamefont {Rogers}},
  \bibinfo {author} {\bibfnamefont {P.}~\bibnamefont {Ricci}}, \ and\ \bibinfo
  {author} {\bibfnamefont {K.~W.}\ \bibnamefont {Gentle}},\ }\bibfield  {title}
  {\enquote {\bibinfo {title} {Plasma transport and turbulence in the helimak:
  Simulation and experiment},}\ }\href {\doibase 10.1063/1.3212591} {\bibfield
  {journal} {\bibinfo  {journal} {Physics of Plasmas}\ }\textbf {\bibinfo
  {volume} {16}},\ \bibinfo {pages} {082510} (\bibinfo {year}
  {2009})}\BibitemShut {NoStop}%
\bibitem [{\citenamefont {Ricci}\ and\ \citenamefont
  {Rogers}(2010)}]{Ricci2010}%
  \BibitemOpen
  \bibfield  {author} {\bibinfo {author} {\bibfnamefont {P.}~\bibnamefont
  {Ricci}}\ and\ \bibinfo {author} {\bibfnamefont {B.~N.}\ \bibnamefont
  {Rogers}},\ }\bibfield  {title} {\enquote {\bibinfo {title} {{Turbulence
  phase space in simple magnetized toroidal plasmas}},}\ }\href {\doibase
  10.1103/PhysRevLett.104.145001} {\bibfield  {journal} {\bibinfo  {journal}
  {Phys. Rev. Lett.}\ }\textbf {\bibinfo {volume} {104}},\ \bibinfo {pages}
  {1--4} (\bibinfo {year} {2010})}\BibitemShut {NoStop}%
\bibitem [{\citenamefont {Li}\ \emph {et~al.}(2011)\citenamefont {Li},
  \citenamefont {Rogers}, \citenamefont {Ricci}, \citenamefont {Gentle},\ and\
  \citenamefont {Bhattacharjee}}]{Li2011}%
  \BibitemOpen
  \bibfield  {author} {\bibinfo {author} {\bibfnamefont {B.}~\bibnamefont
  {Li}}, \bibinfo {author} {\bibfnamefont {B.~N.}\ \bibnamefont {Rogers}},
  \bibinfo {author} {\bibfnamefont {P.}~\bibnamefont {Ricci}}, \bibinfo
  {author} {\bibfnamefont {K.~W.}\ \bibnamefont {Gentle}}, \ and\ \bibinfo
  {author} {\bibfnamefont {A.}~\bibnamefont {Bhattacharjee}},\ }\bibfield
  {title} {\enquote {\bibinfo {title} {Turbulence and bias-induced flows in
  simple magnetized toroidal plasmas},}\ }\href {\doibase
  10.1103/PhysRevE.83.056406} {\bibfield  {journal} {\bibinfo  {journal} {Phys.
  Rev. E}\ }\textbf {\bibinfo {volume} {83}},\ \bibinfo {pages} {056406}
  (\bibinfo {year} {2011})}\BibitemShut {NoStop}%
\bibitem [{\citenamefont {Tamain}\ \emph {et~al.}(2016)\citenamefont {Tamain},
  \citenamefont {Bufferand}, \citenamefont {Ciraolo}, \citenamefont {Colin},
  \citenamefont {Galassi}, \citenamefont {Ghendrih}, \citenamefont
  {Schwander},\ and\ \citenamefont {Serre}}]{Tamain2016}%
  \BibitemOpen
  \bibfield  {author} {\bibinfo {author} {\bibfnamefont {P.}~\bibnamefont
  {Tamain}}, \bibinfo {author} {\bibfnamefont {H.}~\bibnamefont {Bufferand}},
  \bibinfo {author} {\bibfnamefont {G.}~\bibnamefont {Ciraolo}}, \bibinfo
  {author} {\bibfnamefont {C.}~\bibnamefont {Colin}}, \bibinfo {author}
  {\bibfnamefont {D.}~\bibnamefont {Galassi}}, \bibinfo {author} {\bibfnamefont
  {P.}~\bibnamefont {Ghendrih}}, \bibinfo {author} {\bibfnamefont
  {F.}~\bibnamefont {Schwander}}, \ and\ \bibinfo {author} {\bibfnamefont
  {E.}~\bibnamefont {Serre}},\ }\bibfield  {title} {\enquote {\bibinfo {title}
  {{The TOKAM3X code for edge turbulence fluid simulations of tokamak plasmas
  in versatile magnetic geometries}},}\ }\href {\doibase
  10.1016/j.jcp.2016.05.038} {\bibfield  {journal} {\bibinfo  {journal} {J.
  Comput. Phys.}\ }\textbf {\bibinfo {volume} {321}},\ \bibinfo {pages} {1--36}
  (\bibinfo {year} {2016})}\BibitemShut {NoStop}%
\bibitem [{\citenamefont {Halpern}\ \emph {et~al.}(2016)\citenamefont
  {Halpern}, \citenamefont {Ricci}, \citenamefont {Jolliet}, \citenamefont
  {Loizu}, \citenamefont {Morales}, \citenamefont {Mosetto}, \citenamefont
  {Musil}, \citenamefont {Riva}, \citenamefont {Tran},\ and\ \citenamefont
  {Wersal}}]{Halpern2016}%
  \BibitemOpen
  \bibfield  {author} {\bibinfo {author} {\bibfnamefont {F.}~\bibnamefont
  {Halpern}}, \bibinfo {author} {\bibfnamefont {P.}~\bibnamefont {Ricci}},
  \bibinfo {author} {\bibfnamefont {S.}~\bibnamefont {Jolliet}}, \bibinfo
  {author} {\bibfnamefont {J.}~\bibnamefont {Loizu}}, \bibinfo {author}
  {\bibfnamefont {J.}~\bibnamefont {Morales}}, \bibinfo {author} {\bibfnamefont
  {A.}~\bibnamefont {Mosetto}}, \bibinfo {author} {\bibfnamefont
  {F.}~\bibnamefont {Musil}}, \bibinfo {author} {\bibfnamefont
  {F.}~\bibnamefont {Riva}}, \bibinfo {author} {\bibfnamefont {T.}~\bibnamefont
  {Tran}}, \ and\ \bibinfo {author} {\bibfnamefont {C.}~\bibnamefont
  {Wersal}},\ }\bibfield  {title} {\enquote {\bibinfo {title} {{The GBS code
  for tokamak scrape-off layer simulations}},}\ }\href {\doibase
  10.1016/j.jcp.2016.03.040} {\bibfield  {journal} {\bibinfo  {journal} {J.
  Comput. Phys.}\ }\textbf {\bibinfo {volume} {315}},\ \bibinfo {pages}
  {388--408} (\bibinfo {year} {2016})}\BibitemShut {NoStop}%
\bibitem [{\citenamefont {Stegmeir}\ \emph {et~al.}(2018)\citenamefont
  {Stegmeir}, \citenamefont {Coster}, \citenamefont {Ross}, \citenamefont
  {Maj}, \citenamefont {Lackner},\ and\ \citenamefont {Poli}}]{Stegmeir2018}%
  \BibitemOpen
  \bibfield  {author} {\bibinfo {author} {\bibfnamefont {A.}~\bibnamefont
  {Stegmeir}}, \bibinfo {author} {\bibfnamefont {D.}~\bibnamefont {Coster}},
  \bibinfo {author} {\bibfnamefont {A.}~\bibnamefont {Ross}}, \bibinfo {author}
  {\bibfnamefont {O.}~\bibnamefont {Maj}}, \bibinfo {author} {\bibfnamefont
  {K.}~\bibnamefont {Lackner}}, \ and\ \bibinfo {author} {\bibfnamefont
  {E.}~\bibnamefont {Poli}},\ }\bibfield  {title} {\enquote {\bibinfo {title}
  {{GRILLIX: A 3D turbulence code based on the flux-coordinate independent
  approach}},}\ }\href {\doibase 10.1088/1361-6587/aaa373} {\bibfield
  {journal} {\bibinfo  {journal} {Plasma Phys. Control. Fusion}\ }\textbf
  {\bibinfo {volume} {60}} (\bibinfo {year} {2018}),\
  10.1088/1361-6587/aaa373}\BibitemShut {NoStop}%
\bibitem [{\citenamefont {Dudson}\ \emph {et~al.}(2009)\citenamefont {Dudson},
  \citenamefont {Umansky}, \citenamefont {Xu}, \citenamefont {Snyder},\ and\
  \citenamefont {Wilson}}]{Dudson2009}%
  \BibitemOpen
  \bibfield  {author} {\bibinfo {author} {\bibfnamefont {B.~D.}\ \bibnamefont
  {Dudson}}, \bibinfo {author} {\bibfnamefont {M.~V.}\ \bibnamefont {Umansky}},
  \bibinfo {author} {\bibfnamefont {X.~Q.}\ \bibnamefont {Xu}}, \bibinfo
  {author} {\bibfnamefont {P.~B.}\ \bibnamefont {Snyder}}, \ and\ \bibinfo
  {author} {\bibfnamefont {H.~R.}\ \bibnamefont {Wilson}},\ }\bibfield  {title}
  {\enquote {\bibinfo {title} {{BOUT++: A framework for parallel plasma fluid
  simulations}},}\ }\href {\doibase 10.1016/j.cpc.2009.03.008} {\bibfield
  {journal} {\bibinfo  {journal} {Comput. Phys. Commun.}\ }\textbf {\bibinfo
  {volume} {180}},\ \bibinfo {pages} {1467--1480} (\bibinfo {year}
  {2009})}\BibitemShut {NoStop}%
\bibitem [{\citenamefont {Zhu}, \citenamefont {Francisquez},\ and\
  \citenamefont {Rogers}(2018)}]{Zhu2018}%
  \BibitemOpen
  \bibfield  {author} {\bibinfo {author} {\bibfnamefont {B.}~\bibnamefont
  {Zhu}}, \bibinfo {author} {\bibfnamefont {M.}~\bibnamefont {Francisquez}}, \
  and\ \bibinfo {author} {\bibfnamefont {B.~N.}\ \bibnamefont {Rogers}},\
  }\bibfield  {title} {\enquote {\bibinfo {title} {{GDB: A global 3D two-fluid
  model of plasma turbulence and transport in the tokamak edge}},}\ }\href
  {\doibase https://doi.org/10.1016/j.cpc.2018.06.002} {\bibfield  {journal}
  {\bibinfo  {journal} {Computer Physics Communications}\ }\textbf {\bibinfo
  {volume} {232}},\ \bibinfo {pages} {46 -- 58} (\bibinfo {year}
  {2018})}\BibitemShut {NoStop}%
\bibitem [{\citenamefont {Halpern}\ \emph {et~al.}(2015)\citenamefont
  {Halpern}, \citenamefont {Terry}, \citenamefont {Zweben}, \citenamefont
  {LaBombard}, \citenamefont {Podesta},\ and\ \citenamefont
  {Ricci}}]{Halpern2015}%
  \BibitemOpen
  \bibfield  {author} {\bibinfo {author} {\bibfnamefont {F.~D.}\ \bibnamefont
  {Halpern}}, \bibinfo {author} {\bibfnamefont {J.~L.}\ \bibnamefont {Terry}},
  \bibinfo {author} {\bibfnamefont {S.~J.}\ \bibnamefont {Zweben}}, \bibinfo
  {author} {\bibfnamefont {B.}~\bibnamefont {LaBombard}}, \bibinfo {author}
  {\bibfnamefont {M.}~\bibnamefont {Podesta}}, \ and\ \bibinfo {author}
  {\bibfnamefont {P.}~\bibnamefont {Ricci}},\ }\bibfield  {title} {\enquote
  {\bibinfo {title} {{Comparison of 3D flux-driven scrape-off layer turbulence
  simulations with gas-puff imaging of Alcator C-Mod inner-wall limited
  discharges}},}\ }\href {\doibase 10.1088/0741-3335/57/5/054005} {\bibfield
  {journal} {\bibinfo  {journal} {Plasma Phys. Control. Fusion}\ }\textbf
  {\bibinfo {volume} {57}},\ \bibinfo {pages} {054005} (\bibinfo {year}
  {2015})}\BibitemShut {NoStop}%
\bibitem [{\citenamefont {Chen}\ \emph {et~al.}(2017)\citenamefont {Chen},
  \citenamefont {Xu}, \citenamefont {Xia}, \citenamefont {Porkolab},
  \citenamefont {Edlund}, \citenamefont {LaBombard}, \citenamefont {Terry},
  \citenamefont {Hughes}, \citenamefont {Mao}, \citenamefont {Ye},\ and\
  \citenamefont {Wan}}]{Chen2017}%
  \BibitemOpen
  \bibfield  {author} {\bibinfo {author} {\bibfnamefont {B.}~\bibnamefont
  {Chen}}, \bibinfo {author} {\bibfnamefont {X.}~\bibnamefont {Xu}}, \bibinfo
  {author} {\bibfnamefont {T.}~\bibnamefont {Xia}}, \bibinfo {author}
  {\bibfnamefont {M.}~\bibnamefont {Porkolab}}, \bibinfo {author}
  {\bibfnamefont {E.}~\bibnamefont {Edlund}}, \bibinfo {author} {\bibfnamefont
  {B.}~\bibnamefont {LaBombard}}, \bibinfo {author} {\bibfnamefont
  {J.}~\bibnamefont {Terry}}, \bibinfo {author} {\bibfnamefont
  {J.}~\bibnamefont {Hughes}}, \bibinfo {author} {\bibfnamefont
  {S.}~\bibnamefont {Mao}}, \bibinfo {author} {\bibfnamefont {M.}~\bibnamefont
  {Ye}}, \ and\ \bibinfo {author} {\bibfnamefont {Y.}~\bibnamefont {Wan}},\
  }\bibfield  {title} {\enquote {\bibinfo {title} {{Edge turbulence and
  divertor heat flux width simulations of Alcator C-Mod discharges using an
  electromagnetic two-fluid model}},}\ }\href {\doibase
  10.1088/1741-4326/aa7d46} {\bibfield  {journal} {\bibinfo  {journal} {Nuclear
  Fusion}\ }\textbf {\bibinfo {volume} {57}},\ \bibinfo {pages} {116025}
  (\bibinfo {year} {2017})}\BibitemShut {NoStop}%
\bibitem [{\citenamefont {Waltz}\ \emph {et~al.}(2019)\citenamefont {Waltz},
  \citenamefont {Halpern}, \citenamefont {Deng},\ and\ \citenamefont
  {Candy}}]{Waltz2019}%
  \BibitemOpen
  \bibfield  {author} {\bibinfo {author} {\bibfnamefont {R.~E.}\ \bibnamefont
  {Waltz}}, \bibinfo {author} {\bibfnamefont {F.~D.}\ \bibnamefont {Halpern}},
  \bibinfo {author} {\bibfnamefont {Z.}~\bibnamefont {Deng}}, \ and\ \bibinfo
  {author} {\bibfnamefont {J.}~\bibnamefont {Candy}},\ }\href@noop {} {\enquote
  {\bibinfo {title} {Kinetic fluid moments closure for a magnetized plasma with
  collisions},}\ } (\bibinfo {year} {2019}),\ \Eprint
  {http://arxiv.org/abs/arXiv:1901.02429} {arXiv:1901.02429} \BibitemShut
  {NoStop}%
\bibitem [{\citenamefont {Ng}(2018)}]{Ng2018}%
  \BibitemOpen
  \bibfield  {author} {\bibinfo {author} {\bibfnamefont {J.}~\bibnamefont
  {Ng}},\ }\emph {\bibinfo {title} {Fluid closures for the modeling of
  reconnection and instabilities in magnetotail current sheets}},\ \href@noop
  {} {Ph.D. thesis},\ \bibinfo  {school} {Princeton University} (\bibinfo
  {year} {2018})\BibitemShut {NoStop}%
\bibitem [{\citenamefont {Chang}\ \emph {et~al.}(2017)\citenamefont {Chang},
  \citenamefont {Ku}, \citenamefont {Loarte}, \citenamefont {Parail},
  \citenamefont {K{\"{o}}chl}, \citenamefont {Romanelli}, \citenamefont
  {Maingi}, \citenamefont {Ahn}, \citenamefont {Gray}, \citenamefont {Hughes},
  \citenamefont {LaBombard}, \citenamefont {Leonard}, \citenamefont
  {Makowski},\ and\ \citenamefont {Terry}}]{Chang2017}%
  \BibitemOpen
  \bibfield  {author} {\bibinfo {author} {\bibfnamefont {C.~S.}\ \bibnamefont
  {Chang}}, \bibinfo {author} {\bibfnamefont {S.}~\bibnamefont {Ku}}, \bibinfo
  {author} {\bibfnamefont {A.}~\bibnamefont {Loarte}}, \bibinfo {author}
  {\bibfnamefont {V.}~\bibnamefont {Parail}}, \bibinfo {author} {\bibfnamefont
  {F.}~\bibnamefont {K{\"{o}}chl}}, \bibinfo {author} {\bibfnamefont
  {M.}~\bibnamefont {Romanelli}}, \bibinfo {author} {\bibfnamefont
  {R.}~\bibnamefont {Maingi}}, \bibinfo {author} {\bibfnamefont {J.-W.}\
  \bibnamefont {Ahn}}, \bibinfo {author} {\bibfnamefont {T.}~\bibnamefont
  {Gray}}, \bibinfo {author} {\bibfnamefont {J.}~\bibnamefont {Hughes}},
  \bibinfo {author} {\bibfnamefont {B.}~\bibnamefont {LaBombard}}, \bibinfo
  {author} {\bibfnamefont {T.}~\bibnamefont {Leonard}}, \bibinfo {author}
  {\bibfnamefont {M.}~\bibnamefont {Makowski}}, \ and\ \bibinfo {author}
  {\bibfnamefont {J.}~\bibnamefont {Terry}},\ }\bibfield  {title} {\enquote
  {\bibinfo {title} {{Gyrokinetic projection of the divertor heat-flux width
  from present tokamaks to ITER}},}\ }\href {\doibase 10.1088/1741-4326/aa7efb}
  {\bibfield  {journal} {\bibinfo  {journal} {Nucl. Fusion}\ }\textbf {\bibinfo
  {volume} {57}},\ \bibinfo {pages} {116023} (\bibinfo {year}
  {2017})}\BibitemShut {NoStop}%
\bibitem [{\citenamefont {Pan}\ \emph {et~al.}(2018)\citenamefont {Pan},
  \citenamefont {Told}, \citenamefont {Shi}, \citenamefont {Hammett},\ and\
  \citenamefont {Jenko}}]{Pan2018}%
  \BibitemOpen
  \bibfield  {author} {\bibinfo {author} {\bibfnamefont {Q.}~\bibnamefont
  {Pan}}, \bibinfo {author} {\bibfnamefont {D.}~\bibnamefont {Told}}, \bibinfo
  {author} {\bibfnamefont {E.}~\bibnamefont {Shi}}, \bibinfo {author}
  {\bibfnamefont {G.~W.}\ \bibnamefont {Hammett}}, \ and\ \bibinfo {author}
  {\bibfnamefont {F.}~\bibnamefont {Jenko}},\ }\bibfield  {title} {\enquote
  {\bibinfo {title} {{Full-$f$ version of GENE for turbulence in
  open-field-line systems}},}\ }\href {\doibase 10.1063/1.5008895} {\bibfield
  {journal} {\bibinfo  {journal} {Phys. Plasmas}\ }\textbf {\bibinfo {volume}
  {25}},\ \bibinfo {pages} {1--11} (\bibinfo {year} {2018})}\BibitemShut
  {NoStop}%
\bibitem [{\citenamefont {Grandgirard}\ \emph {et~al.}(2016)\citenamefont
  {Grandgirard}, \citenamefont {Abiteboul}, \citenamefont {Bigot},
  \citenamefont {Cartier-michaud},\ and\ \citenamefont
  {Crouseilles}}]{Grandgirard2016}%
  \BibitemOpen
  \bibfield  {author} {\bibinfo {author} {\bibfnamefont {V.}~\bibnamefont
  {Grandgirard}}, \bibinfo {author} {\bibfnamefont {J.}~\bibnamefont
  {Abiteboul}}, \bibinfo {author} {\bibfnamefont {J.}~\bibnamefont {Bigot}},
  \bibinfo {author} {\bibfnamefont {T.}~\bibnamefont {Cartier-michaud}}, \ and\
  \bibinfo {author} {\bibfnamefont {N.}~\bibnamefont {Crouseilles}},\
  }\bibfield  {title} {\enquote {\bibinfo {title} {{A 5D gyrokinetic full-$f$
  global semi-Lagrangian code for flux-driven ion turbulence simulations}},}\
  }\href {\doibase 10.1016/j.cpc.2016.05.007} {\bibfield  {journal} {\bibinfo
  {journal} {Comput. Phys. Commun.}\ }\textbf {\bibinfo {volume} {207}},\
  \bibinfo {pages} {35--68} (\bibinfo {year} {2016})}\BibitemShut {NoStop}%
\bibitem [{\citenamefont {Ch{\^{o}}n{\'{e}}}\ \emph {et~al.}(2018)\citenamefont
  {Ch{\^{o}}n{\'{e}}}, \citenamefont {Kiviniemi}, \citenamefont {Leerink},
  \citenamefont {Niskala},\ and\ \citenamefont {Rochford}}]{Chone2018}%
  \BibitemOpen
  \bibfield  {author} {\bibinfo {author} {\bibfnamefont {L.}~\bibnamefont
  {Ch{\^{o}}n{\'{e}}}}, \bibinfo {author} {\bibfnamefont {T.~P.}\ \bibnamefont
  {Kiviniemi}}, \bibinfo {author} {\bibfnamefont {S.}~\bibnamefont {Leerink}},
  \bibinfo {author} {\bibfnamefont {P.}~\bibnamefont {Niskala}}, \ and\
  \bibinfo {author} {\bibfnamefont {R.}~\bibnamefont {Rochford}},\ }\bibfield
  {title} {\enquote {\bibinfo {title} {{Improved boundary condition for
  full-$f$ gyrokinetic simulations of circular-limited tokamak plasmas in
  ELMFIRE}},}\ }\href {\doibase 10.1002/ctpp.201700185} {\bibfield  {journal}
  {\bibinfo  {journal} {Contrib. to Plasma Phys.}\ }\textbf {\bibinfo {volume}
  {58}},\ \bibinfo {pages} {534--539} (\bibinfo {year} {2018})}\BibitemShut
  {NoStop}%
\bibitem [{\citenamefont {Boesl}\ \emph {et~al.}(2019)\citenamefont {Boesl},
  \citenamefont {Bergmann}, \citenamefont {Bottino}, \citenamefont {Coster},
  \citenamefont {Lanti}, \citenamefont {Ohana},\ and\ \citenamefont
  {Jenko}}]{Boesl2019}%
  \BibitemOpen
  \bibfield  {author} {\bibinfo {author} {\bibfnamefont {M.~H.}\ \bibnamefont
  {Boesl}}, \bibinfo {author} {\bibfnamefont {A.}~\bibnamefont {Bergmann}},
  \bibinfo {author} {\bibfnamefont {A.}~\bibnamefont {Bottino}}, \bibinfo
  {author} {\bibfnamefont {D.}~\bibnamefont {Coster}}, \bibinfo {author}
  {\bibfnamefont {E.}~\bibnamefont {Lanti}}, \bibinfo {author} {\bibfnamefont
  {N.}~\bibnamefont {Ohana}}, \ and\ \bibinfo {author} {\bibfnamefont
  {F.}~\bibnamefont {Jenko}},\ }\href@noop {} {\enquote {\bibinfo {title}
  {{Gyrokinetic full-$f$ particle-in-cell simulations on open field lines with
  PICLS}},}\ } (\bibinfo {year} {2019}),\ \Eprint
  {http://arxiv.org/abs/arXiv:1908.00318} {arXiv:1908.00318} \BibitemShut
  {NoStop}%
\bibitem [{\citenamefont {Dorf}\ \emph {et~al.}(2016)\citenamefont {Dorf},
  \citenamefont {Dorr}, \citenamefont {Hittinger}, \citenamefont {Cohen},\ and\
  \citenamefont {Rognlien}}]{Dorf2016}%
  \BibitemOpen
  \bibfield  {author} {\bibinfo {author} {\bibfnamefont {M.~A.}\ \bibnamefont
  {Dorf}}, \bibinfo {author} {\bibfnamefont {M.~R.}\ \bibnamefont {Dorr}},
  \bibinfo {author} {\bibfnamefont {J.~A.}\ \bibnamefont {Hittinger}}, \bibinfo
  {author} {\bibfnamefont {R.~H.}\ \bibnamefont {Cohen}}, \ and\ \bibinfo
  {author} {\bibfnamefont {T.~D.}\ \bibnamefont {Rognlien}},\ }\bibfield
  {title} {\enquote {\bibinfo {title} {{Continuum kinetic modeling of the
  tokamak plasma edge}},}\ }\href {\doibase 10.1063/1.4943106} {\bibfield
  {journal} {\bibinfo  {journal} {Phys. Plasmas}\ }\textbf {\bibinfo {volume}
  {23}},\ \bibinfo {pages} {056102} (\bibinfo {year} {2016})}\BibitemShut
  {NoStop}%
\bibitem [{\citenamefont {Shi}\ \emph {et~al.}(2017)\citenamefont {Shi},
  \citenamefont {Hammett}, \citenamefont {Stoltzfus-Dueck},\ and\ \citenamefont
  {Hakim}}]{Shi2017}%
  \BibitemOpen
  \bibfield  {author} {\bibinfo {author} {\bibfnamefont {E.~L.}\ \bibnamefont
  {Shi}}, \bibinfo {author} {\bibfnamefont {G.~W.}\ \bibnamefont {Hammett}},
  \bibinfo {author} {\bibfnamefont {T.}~\bibnamefont {Stoltzfus-Dueck}}, \ and\
  \bibinfo {author} {\bibfnamefont {A.}~\bibnamefont {Hakim}},\ }\bibfield
  {title} {\enquote {\bibinfo {title} {Gyrokinetic continuum simulation of
  turbulence in a straight open-field-line plasma},}\ }\href {\doibase
  10.1017/S002237781700037X} {\bibfield  {journal} {\bibinfo  {journal}
  {Journal of Plasma Physics}\ }\textbf {\bibinfo {volume} {83}},\ \bibinfo
  {pages} {905830304} (\bibinfo {year} {2017})}\BibitemShut {NoStop}%
\bibitem [{\citenamefont {Shi}\ \emph {et~al.}(2019)\citenamefont {Shi},
  \citenamefont {Hammett}, \citenamefont {Stoltzfus-Dueck},\ and\ \citenamefont
  {Hakim}}]{Shi2019}%
  \BibitemOpen
  \bibfield  {author} {\bibinfo {author} {\bibfnamefont {E.~L.}\ \bibnamefont
  {Shi}}, \bibinfo {author} {\bibfnamefont {G.~W.}\ \bibnamefont {Hammett}},
  \bibinfo {author} {\bibfnamefont {T.}~\bibnamefont {Stoltzfus-Dueck}}, \ and\
  \bibinfo {author} {\bibfnamefont {A.}~\bibnamefont {Hakim}},\ }\bibfield
  {title} {\enquote {\bibinfo {title} {Full-$f$ gyrokinetic simulation of
  turbulence in a helical open-field-line plasma},}\ }\href {\doibase
  10.1063/1.5074179} {\bibfield  {journal} {\bibinfo  {journal} {Physics of
  Plasmas}\ }\textbf {\bibinfo {volume} {26}},\ \bibinfo {pages} {012307}
  (\bibinfo {year} {2019})}\BibitemShut {NoStop}%
\bibitem [{\citenamefont {Bernard}\ \emph {et~al.}(2019)\citenamefont
  {Bernard}, \citenamefont {Shi}, \citenamefont {Gentle}, \citenamefont
  {Hakim}, \citenamefont {Hammett}, \citenamefont {Stoltzfus-Dueck},\ and\
  \citenamefont {Taylor}}]{Bernard2019}%
  \BibitemOpen
  \bibfield  {author} {\bibinfo {author} {\bibfnamefont {T.~N.}\ \bibnamefont
  {Bernard}}, \bibinfo {author} {\bibfnamefont {E.~L.}\ \bibnamefont {Shi}},
  \bibinfo {author} {\bibfnamefont {K.~W.}\ \bibnamefont {Gentle}}, \bibinfo
  {author} {\bibfnamefont {A.}~\bibnamefont {Hakim}}, \bibinfo {author}
  {\bibfnamefont {G.~W.}\ \bibnamefont {Hammett}}, \bibinfo {author}
  {\bibfnamefont {T.}~\bibnamefont {Stoltzfus-Dueck}}, \ and\ \bibinfo {author}
  {\bibfnamefont {E.~I.}\ \bibnamefont {Taylor}},\ }\bibfield  {title}
  {\enquote {\bibinfo {title} {Gyrokinetic continuum simulations of plasma
  turbulence in the texas helimak},}\ }\href {\doibase 10.1063/1.5085457}
  {\bibfield  {journal} {\bibinfo  {journal} {Physics of Plasmas}\ }\textbf
  {\bibinfo {volume} {26}},\ \bibinfo {pages} {042301} (\bibinfo {year}
  {2019})}\BibitemShut {NoStop}%
\bibitem [{\citenamefont {Shi}(2017)}]{Shi2017a}%
  \BibitemOpen
  \bibfield  {author} {\bibinfo {author} {\bibfnamefont {E.~L.}\ \bibnamefont
  {Shi}},\ }\emph {\bibinfo {title} {Gyrokinetic continuum simulation of
  turbulence in open--field--line plasmas}},\ \href@noop {} {Ph.D. thesis},\
  \bibinfo  {school} {Princeton University} (\bibinfo {year}
  {2017})\BibitemShut {NoStop}%
\bibitem [{\citenamefont {Hakim}\ \emph {et~al.}(2019)\citenamefont {Hakim},
  \citenamefont {Hammett}, \citenamefont {Shi},\ and\ \citenamefont
  {Mandell}}]{Hakim2019}%
  \BibitemOpen
  \bibfield  {author} {\bibinfo {author} {\bibfnamefont {A.}~\bibnamefont
  {Hakim}}, \bibinfo {author} {\bibfnamefont {G.}~\bibnamefont {Hammett}},
  \bibinfo {author} {\bibfnamefont {E.}~\bibnamefont {Shi}}, \ and\ \bibinfo
  {author} {\bibfnamefont {N.}~\bibnamefont {Mandell}},\ }\href@noop {}
  {\enquote {\bibinfo {title} {Discontinuous galerkin schemes for a class of
  hamiltonian evolution equations with applications to plasma fluid and kinetic
  problems},}\ } (\bibinfo {year} {2019}),\ \Eprint
  {http://arxiv.org/abs/arXiv:1908.01814} {arXiv:1908.01814} \BibitemShut
  {NoStop}%
\bibitem [{\citenamefont {Geraldini}, \citenamefont {Parra},\ and\
  \citenamefont {Militello}(2017)}]{Geraldini2017}%
  \BibitemOpen
  \bibfield  {author} {\bibinfo {author} {\bibfnamefont {A.}~\bibnamefont
  {Geraldini}}, \bibinfo {author} {\bibfnamefont {F.~I.}\ \bibnamefont
  {Parra}}, \ and\ \bibinfo {author} {\bibfnamefont {F.}~\bibnamefont
  {Militello}},\ }\bibfield  {title} {\enquote {\bibinfo {title} {Gyrokinetic
  treatment of a grazing angle magnetic presheath},}\ }\href {\doibase
  10.1088/1361-6587/59/2/025015} {\bibfield  {journal} {\bibinfo  {journal}
  {Plasma Physics and Controlled Fusion}\ }\textbf {\bibinfo {volume} {59}},\
  \bibinfo {pages} {025015} (\bibinfo {year} {2017})}\BibitemShut {NoStop}%
\bibitem [{\citenamefont {Madsen}\ \emph {et~al.}(2016)\citenamefont {Madsen},
  \citenamefont {Naulin}, \citenamefont {Nielsen},\ and\ \citenamefont
  {Rasmussen}}]{Madsen2016}%
  \BibitemOpen
  \bibfield  {author} {\bibinfo {author} {\bibfnamefont {J.}~\bibnamefont
  {Madsen}}, \bibinfo {author} {\bibfnamefont {V.}~\bibnamefont {Naulin}},
  \bibinfo {author} {\bibfnamefont {A.~H.}\ \bibnamefont {Nielsen}}, \ and\
  \bibinfo {author} {\bibfnamefont {J.~J.}\ \bibnamefont {Rasmussen}},\
  }\bibfield  {title} {\enquote {\bibinfo {title} {{Collisional transport
  across the magnetic field in drift-fluid models}},}\ }\href {\doibase
  10.1063/1.4943199} {\bibfield  {journal} {\bibinfo  {journal} {Phys.
  Plasmas}\ }\textbf {\bibinfo {volume} {23}} (\bibinfo {year} {2016}),\
  10.1063/1.4943199}\BibitemShut {NoStop}%
\bibitem [{\citenamefont {Francisquez}(2018)}]{Francisquez2018a}%
  \BibitemOpen
  \bibfield  {author} {\bibinfo {author} {\bibfnamefont {M.}~\bibnamefont
  {Francisquez}},\ }\emph {\bibinfo {title} {{Global Braginskii modeling of
  magnetically confined boundary plasmas}}},\ \href@noop {} {Ph.D. thesis},\
  \bibinfo  {school} {Dartmouth College}, \bibinfo {address} {{Hanover, NH
  03755}} (\bibinfo {year} {2018})\BibitemShut {NoStop}%
\bibitem [{\citenamefont {Huba}(2013)}]{Huba2013}%
  \BibitemOpen
  \bibfield  {author} {\bibinfo {author} {\bibfnamefont {J.~D.}\ \bibnamefont
  {Huba}},\ }\href {http://wwwppd.nrl.navy.mil/nrlformulary/} {\emph {\bibinfo
  {title} {Plasma Physics}}}\ (\bibinfo  {publisher} {Naval Research
  Laboratory},\ \bibinfo {address} {Washington, DC},\ \bibinfo {year} {2013})\
  pp.\ \bibinfo {pages} {1--71}\BibitemShut {NoStop}%
\bibitem [{\citenamefont {Francisquez}, \citenamefont {Zhu},\ and\
  \citenamefont {Rogers}(2019)}]{Francisquez2019}%
  \BibitemOpen
  \bibfield  {author} {\bibinfo {author} {\bibfnamefont {M.}~\bibnamefont
  {Francisquez}}, \bibinfo {author} {\bibfnamefont {B.}~\bibnamefont {Zhu}}, \
  and\ \bibinfo {author} {\bibfnamefont {B.~N.}\ \bibnamefont {Rogers}},\
  }\bibfield  {title} {\enquote {\bibinfo {title} {{Multigrid treatment of
  implicit continuum diffusion}},}\ }\href {\doibase 10.1016/j.cpc.2018.10.022}
  {\bibfield  {journal} {\bibinfo  {journal} {Comput. Phys. Commun.}\ }\textbf
  {\bibinfo {volume} {236}},\ \bibinfo {pages} {104--117} (\bibinfo {year}
  {2019})}\BibitemShut {NoStop}%
\bibitem [{\citenamefont {Loizu}\ \emph {et~al.}(2012)\citenamefont {Loizu},
  \citenamefont {Ricci}, \citenamefont {Halpern},\ and\ \citenamefont
  {Jolliet}}]{Loizu2012}%
  \BibitemOpen
  \bibfield  {author} {\bibinfo {author} {\bibfnamefont {J.}~\bibnamefont
  {Loizu}}, \bibinfo {author} {\bibfnamefont {P.}~\bibnamefont {Ricci}},
  \bibinfo {author} {\bibfnamefont {F.~D.}\ \bibnamefont {Halpern}}, \ and\
  \bibinfo {author} {\bibfnamefont {S.}~\bibnamefont {Jolliet}},\ }\bibfield
  {title} {\enquote {\bibinfo {title} {Boundary conditions for plasma fluid
  models at the magnetic presheath entrance},}\ }\href {\doibase
  10.1063/1.4771573} {\bibfield  {journal} {\bibinfo  {journal} {Physics of
  Plasmas}\ }\textbf {\bibinfo {volume} {19}},\ \bibinfo {pages} {122307}
  (\bibinfo {year} {2012})},\ \Eprint
  {http://arxiv.org/abs/https://doi.org/10.1063/1.4771573}
  {https://doi.org/10.1063/1.4771573} \BibitemShut {NoStop}%
\bibitem [{\citenamefont {Stegmeir}\ \emph {et~al.}(2019)\citenamefont
  {Stegmeir}, \citenamefont {Ross}, \citenamefont {Body}, \citenamefont
  {Francisquez}, \citenamefont {Zholobenko}, \citenamefont {Coster},
  \citenamefont {Maj}, \citenamefont {Manz}, \citenamefont {Jenko},
  \citenamefont {Rogers},\ and\ \citenamefont {Kang}}]{Stegmeir2019}%
  \BibitemOpen
  \bibfield  {author} {\bibinfo {author} {\bibfnamefont {A.}~\bibnamefont
  {Stegmeir}}, \bibinfo {author} {\bibfnamefont {A.}~\bibnamefont {Ross}},
  \bibinfo {author} {\bibfnamefont {T.}~\bibnamefont {Body}}, \bibinfo {author}
  {\bibfnamefont {M.}~\bibnamefont {Francisquez}}, \bibinfo {author}
  {\bibfnamefont {W.}~\bibnamefont {Zholobenko}}, \bibinfo {author}
  {\bibfnamefont {D.}~\bibnamefont {Coster}}, \bibinfo {author} {\bibfnamefont
  {O.}~\bibnamefont {Maj}}, \bibinfo {author} {\bibfnamefont {P.}~\bibnamefont
  {Manz}}, \bibinfo {author} {\bibfnamefont {F.}~\bibnamefont {Jenko}},
  \bibinfo {author} {\bibfnamefont {B.~N.}\ \bibnamefont {Rogers}}, \ and\
  \bibinfo {author} {\bibfnamefont {K.~S.}\ \bibnamefont {Kang}},\ }\bibfield
  {title} {\enquote {\bibinfo {title} {{Global turbulence simulations of the
  tokamak edge region with GRILLIX}},}\ }\href {\doibase 10.1063/1.5089864}
  {\bibfield  {journal} {\bibinfo  {journal} {Phys. Plasmas}\ }\textbf
  {\bibinfo {volume} {26}} (\bibinfo {year} {2019}),\
  10.1063/1.5089864}\BibitemShut {NoStop}%
\bibitem [{\citenamefont {Smith}\ and\ \citenamefont
  {Hammett}(1997)}]{Smith1997}%
  \BibitemOpen
  \bibfield  {author} {\bibinfo {author} {\bibfnamefont {S.~A.}\ \bibnamefont
  {Smith}}\ and\ \bibinfo {author} {\bibfnamefont {G.~W.}\ \bibnamefont
  {Hammett}},\ }\bibfield  {title} {\enquote {\bibinfo {title} {{Eddy viscosity
  and hyperviscosity in spectral simulations of 2D drift wave turbulence}},}\
  }\href {\doibase 10.1063/1.872210} {\bibfield  {journal} {\bibinfo  {journal}
  {Phys. Plasmas}\ }\textbf {\bibinfo {volume} {4}},\ \bibinfo {pages} {978}
  (\bibinfo {year} {1997})}\BibitemShut {NoStop}%
\bibitem [{\citenamefont {Xia}, \citenamefont {Xu},\ and\ \citenamefont
  {Xi}(2013)}]{Xia2013}%
  \BibitemOpen
  \bibfield  {author} {\bibinfo {author} {\bibfnamefont {T.}~\bibnamefont
  {Xia}}, \bibinfo {author} {\bibfnamefont {X.}~\bibnamefont {Xu}}, \ and\
  \bibinfo {author} {\bibfnamefont {P.}~\bibnamefont {Xi}},\ }\bibfield
  {title} {\enquote {\bibinfo {title} {Six-field two-fluid simulations of
  peeling{\textendash}ballooning modes using {BOUT++}},}\ }\href {\doibase
  10.1088/0029-5515/53/7/073009} {\bibfield  {journal} {\bibinfo  {journal}
  {Nuclear Fusion}\ }\textbf {\bibinfo {volume} {53}},\ \bibinfo {pages}
  {073009} (\bibinfo {year} {2013})}\BibitemShut {NoStop}%
\bibitem [{\citenamefont {Halpern}\ \emph {et~al.}(2014)\citenamefont
  {Halpern}, \citenamefont {Ricci}, \citenamefont {Jolliet}, \citenamefont
  {Loizu},\ and\ \citenamefont {Mosetto}}]{Halpern2014}%
  \BibitemOpen
  \bibfield  {author} {\bibinfo {author} {\bibfnamefont {F.}~\bibnamefont
  {Halpern}}, \bibinfo {author} {\bibfnamefont {P.}~\bibnamefont {Ricci}},
  \bibinfo {author} {\bibfnamefont {S.}~\bibnamefont {Jolliet}}, \bibinfo
  {author} {\bibfnamefont {J.}~\bibnamefont {Loizu}}, \ and\ \bibinfo {author}
  {\bibfnamefont {A.}~\bibnamefont {Mosetto}},\ }\bibfield  {title} {\enquote
  {\bibinfo {title} {Theory of the scrape-off layer width in inner-wall limited
  tokamak plasmas},}\ }\href {\doibase 10.1088/0029-5515/54/4/043003}
  {\bibfield  {journal} {\bibinfo  {journal} {Nuclear Fusion}\ }\textbf
  {\bibinfo {volume} {54}},\ \bibinfo {pages} {043003} (\bibinfo {year}
  {2014})}\BibitemShut {NoStop}%
\bibitem [{\citenamefont {Francisquez}, \citenamefont {Zhu},\ and\
  \citenamefont {Rogers}(2017)}]{Francisquez2017}%
  \BibitemOpen
  \bibfield  {author} {\bibinfo {author} {\bibfnamefont {M.}~\bibnamefont
  {Francisquez}}, \bibinfo {author} {\bibfnamefont {B.}~\bibnamefont {Zhu}}, \
  and\ \bibinfo {author} {\bibfnamefont {B.}~\bibnamefont {Rogers}},\
  }\bibfield  {title} {\enquote {\bibinfo {title} {{Global 3D Braginskii
  simulations of the tokamak edge region of {IWL} discharges}},}\ }\href
  {\doibase 10.1088/1741-4326/aa7f23} {\bibfield  {journal} {\bibinfo
  {journal} {Nuclear Fusion}\ }\textbf {\bibinfo {volume} {57}},\ \bibinfo
  {pages} {116049} (\bibinfo {year} {2017})}\BibitemShut {NoStop}%
\bibitem [{\citenamefont {Dudson}\ and\ \citenamefont
  {Leddy}(2017)}]{Dudson2017}%
  \BibitemOpen
  \bibfield  {author} {\bibinfo {author} {\bibfnamefont {B.~D.}\ \bibnamefont
  {Dudson}}\ and\ \bibinfo {author} {\bibfnamefont {J.}~\bibnamefont {Leddy}},\
  }\bibfield  {title} {\enquote {\bibinfo {title} {Hermes: global plasma edge
  fluid turbulence simulations},}\ }\href {\doibase 10.1088/1361-6587/aa63d2}
  {\bibfield  {journal} {\bibinfo  {journal} {Plasma Physics and Controlled
  Fusion}\ }\textbf {\bibinfo {volume} {59}},\ \bibinfo {pages} {054010}
  (\bibinfo {year} {2017})}\BibitemShut {NoStop}%
\bibitem [{\citenamefont {D'Ippolito}, \citenamefont {Myra},\ and\
  \citenamefont {Zweben}(2011)}]{d2011convective}%
  \BibitemOpen
  \bibfield  {author} {\bibinfo {author} {\bibfnamefont {D.}~\bibnamefont
  {D'Ippolito}}, \bibinfo {author} {\bibfnamefont {J.}~\bibnamefont {Myra}}, \
  and\ \bibinfo {author} {\bibfnamefont {S.}~\bibnamefont {Zweben}},\
  }\bibfield  {title} {\enquote {\bibinfo {title} {{Convective transport by
  intermittent blob-filaments: Comparison of theory and experiment}},}\ }\href
  {\doibase 10.1063/1.3594609} {\bibfield  {journal} {\bibinfo  {journal}
  {Phys. Plasmas}\ }\textbf {\bibinfo {volume} {18}},\ \bibinfo {pages}
  {060501} (\bibinfo {year} {2011})}\BibitemShut {NoStop}%
\bibitem [{\citenamefont {Anderson}\ and\ \citenamefont
  {O'Neil}(2007)}]{Anderson2007}%
  \BibitemOpen
  \bibfield  {author} {\bibinfo {author} {\bibfnamefont {M.~W.}\ \bibnamefont
  {Anderson}}\ and\ \bibinfo {author} {\bibfnamefont {T.~M.}\ \bibnamefont
  {O'Neil}},\ }\bibfield  {title} {\enquote {\bibinfo {title} {{Eigenfunctions
  and eigenvalues of the Dougherty collision operator}},}\ }\href {\doibase
  10.1063/1.2727463} {\bibfield  {journal} {\bibinfo  {journal} {Phys.
  Plasmas}\ }\textbf {\bibinfo {volume} {14}},\ \bibinfo {pages} {052103}
  (\bibinfo {year} {2007})}\BibitemShut {NoStop}%
\bibitem [{\citenamefont {Pan}\ and\ \citenamefont {Ernst}(2019)}]{Pan2019}%
  \BibitemOpen
  \bibfield  {author} {\bibinfo {author} {\bibfnamefont {Q.}~\bibnamefont
  {Pan}}\ and\ \bibinfo {author} {\bibfnamefont {D.~R.}\ \bibnamefont
  {Ernst}},\ }\bibfield  {title} {\enquote {\bibinfo {title} {{Gyrokinetic
  Landau collision operator in conservative form}},}\ }\href {\doibase
  10.1103/PhysRevE.99.023201} {\bibfield  {journal} {\bibinfo  {journal} {Phys.
  Rev. E}\ }\textbf {\bibinfo {volume} {99}},\ \bibinfo {pages} {023201}
  (\bibinfo {year} {2019})}\BibitemShut {NoStop}%
\bibitem [{\citenamefont {Sugama}\ \emph {et~al.}(2019)\citenamefont {Sugama},
  \citenamefont {Matsuoka}, \citenamefont {Satake}, \citenamefont {Nunami},\
  and\ \citenamefont {Watanabe}}]{Sugama2019}%
  \BibitemOpen
  \bibfield  {author} {\bibinfo {author} {\bibfnamefont {H.}~\bibnamefont
  {Sugama}}, \bibinfo {author} {\bibfnamefont {S.}~\bibnamefont {Matsuoka}},
  \bibinfo {author} {\bibfnamefont {S.}~\bibnamefont {Satake}}, \bibinfo
  {author} {\bibfnamefont {M.}~\bibnamefont {Nunami}}, \ and\ \bibinfo {author}
  {\bibfnamefont {T.}~\bibnamefont {Watanabe}},\ }\href@noop {} {\enquote
  {\bibinfo {title} {Improved linearized model collision operator for the
  highly collisional regime},}\ } (\bibinfo {year} {2019}),\ \Eprint
  {http://arxiv.org/abs/arXiv:1906.07427} {arXiv:1906.07427} \BibitemShut
  {NoStop}%
\bibitem [{\citenamefont {Jorge}, \citenamefont {Frei},\ and\ \citenamefont
  {Ricci}(2019)}]{Jorge2019}%
  \BibitemOpen
  \bibfield  {author} {\bibinfo {author} {\bibfnamefont {R.}~\bibnamefont
  {Jorge}}, \bibinfo {author} {\bibfnamefont {B.~J.}\ \bibnamefont {Frei}}, \
  and\ \bibinfo {author} {\bibfnamefont {P.}~\bibnamefont {Ricci}},\
  }\href@noop {} {\enquote {\bibinfo {title} {{Non-Linear Gyrokinetic Coulomb
  Collision Operator}},}\ } (\bibinfo {year} {2019}),\ \Eprint
  {http://arxiv.org/abs/arXiv:1906.03252} {arXiv:1906.03252} \BibitemShut
  {NoStop}%
\bibitem [{\citenamefont {Halpern}\ and\ \citenamefont
  {Waltz}(2018)}]{Halpern2018}%
  \BibitemOpen
  \bibfield  {author} {\bibinfo {author} {\bibfnamefont {F.~D.}\ \bibnamefont
  {Halpern}}\ and\ \bibinfo {author} {\bibfnamefont {R.~E.}\ \bibnamefont
  {Waltz}},\ }\bibfield  {title} {\enquote {\bibinfo {title} {Anti-symmetric
  plasma moment equations with conservative discrete counterparts},}\ }\href
  {\doibase 10.1063/1.5038110} {\bibfield  {journal} {\bibinfo  {journal}
  {Physics of Plasmas}\ }\textbf {\bibinfo {volume} {25}},\ \bibinfo {pages}
  {060703} (\bibinfo {year} {2018})}\BibitemShut {NoStop}%
\bibitem [{\citenamefont {Dejarnac}\ \emph {et~al.}(2015)\citenamefont
  {Dejarnac}, \citenamefont {Stangeby}, \citenamefont {Goldston}, \citenamefont
  {Gauthier}, \citenamefont {Horacek}, \citenamefont {Hron}, \citenamefont
  {Kocan}, \citenamefont {Komm}, \citenamefont {Panek}, \citenamefont {Pitts},\
  and\ \citenamefont {Vondracek}}]{Dejarnac2015}%
  \BibitemOpen
  \bibfield  {author} {\bibinfo {author} {\bibfnamefont {R.}~\bibnamefont
  {Dejarnac}}, \bibinfo {author} {\bibfnamefont {P.}~\bibnamefont {Stangeby}},
  \bibinfo {author} {\bibfnamefont {R.}~\bibnamefont {Goldston}}, \bibinfo
  {author} {\bibfnamefont {E.}~\bibnamefont {Gauthier}}, \bibinfo {author}
  {\bibfnamefont {J.}~\bibnamefont {Horacek}}, \bibinfo {author} {\bibfnamefont
  {M.}~\bibnamefont {Hron}}, \bibinfo {author} {\bibfnamefont {M.}~\bibnamefont
  {Kocan}}, \bibinfo {author} {\bibfnamefont {M.}~\bibnamefont {Komm}},
  \bibinfo {author} {\bibfnamefont {R.}~\bibnamefont {Panek}}, \bibinfo
  {author} {\bibfnamefont {R.}~\bibnamefont {Pitts}}, \ and\ \bibinfo {author}
  {\bibfnamefont {P.}~\bibnamefont {Vondracek}},\ }\bibfield  {title} {\enquote
  {\bibinfo {title} {Understanding narrow sol power flux component in compass
  limiter plasmas by use of langmuir probes},}\ }\href {\doibase
  https://doi.org/10.1016/j.jnucmat.2014.12.100} {\bibfield  {journal}
  {\bibinfo  {journal} {Journal of Nuclear Materials}\ }\textbf {\bibinfo
  {volume} {463}},\ \bibinfo {pages} {381 -- 384} (\bibinfo {year}
  {2015})}\BibitemShut {NoStop}%
\bibitem [{\citenamefont {Ross}(2018)}]{Ross2018}%
  \BibitemOpen
  \bibfield  {author} {\bibinfo {author} {\bibfnamefont {A.}~\bibnamefont
  {Ross}},\ }\emph {\bibinfo {title} {{Extension of GRILLIX: Towards a global
  fluid turbulence code for realistic magnetic geometries}}},\ \href@noop {}
  {Ph.D. thesis},\ \bibinfo  {school} {Technische Universit{\"{a}}t
  M{\"{u}}nchen} (\bibinfo {year} {2018})\BibitemShut {NoStop}%
\end{thebibliography}%

\end{document}